\documentclass[iop,apj,tighten,numberedappendix,twocolappendix]{emulateapj}
\usepackage{natbib}
\usepackage{amsmath}

\newcommand{\figref}[1]{Figure~\ref{fig:#1}}
\newcommand{\tabref}[1]{Table~\ref{tab:#1}}
\newcommand{\secref}[1]{\S\ref{sec:#1}}
\newcommand{\secrefword}[1]{Section \ref{sec:#1}}

\newcommand{\subsecref}[1]{\S\ref{subsec:#1}}
\newcommand{\appref}[1]{Appendix~\ref{app:#1}}
\newcommand{\eqnref}[1]{Equation~\ref{eqn:#1}}

\shorttitle{Mergers in Galaxy Groups. II}
\shortauthors{Taranu, Dubinski, \& Yee}

\begin{document}

\title{Mergers in Galaxy Groups. II. The Fundamental Plane of Elliptical Galaxies}
\author{Dan Taranu, John Dubinski, and H.K.C. Yee}
\affil{Department of Astronomy and Astrophysics, University of Toronto, 50 St. George Street, Toronto, Ontario, Canada, M5S 3H4}
\email{taranu@astro.utoronto.ca}

\begin{abstract}
Observations consistently show that elliptical galaxies follow a tight ``fundamental plane'' scaling relation between size, mean surface brightness and velocity dispersion, with the form $R \propto \sigma^{a}\mu^{b}$. This relation not only has very small ($<$ 0.05 dex) intrinsic scatter, but also has significantly different coefficients from the expected virial scaling (a ``tilt''). We analyze hundreds of simulations of elliptical galaxies formed from mergers of spiral galaxies in groups to determine if the fundamental plane can emerge from multiple, mostly minor and hierarchical collisionless mergers. We find that these simulated ellipticals lie on a similar fundamental plane with $a \approx 1.7$ and $b \approx 0.3$. The scatter about this plane is not larger than observed, while the tilt is in the correct sense, although $a$ is larger than for typical observations. This supports the idea that collisionless mergers can contribute significantly to the tilt of the fundamental plane. The tilt is mainly driven by a mass-dependent dark matter fraction, such that more massive galaxies have larger dark matter fractions within $R_{e}$. We further discuss the origin of this mass-dependent dark matter fraction and its compatibility with strong lensing observations, as well as the links between the fundamental plane, dynamical masses and the virial theorem.
\end{abstract}


\section{Introduction}
\label{sec:introduction}

Three simple but fundamental properties of galaxies are their sizes, luminosities and velocity dispersions. Elliptical galaxies show some of the strongest correlations between these properties, with small or even negligible internal scatter. Explaining the origin of these scaling relations is a long-standing challenge in elliptical formation theory.

Two of the key scaling relations of elliptical galaxies were first noted over 35 years ago. The first of these is the Faber-Jackson (FJ) relation \citep{FabJac76} between velocity dispersion $\sigma$ and luminosity $L$, typically observed to be $L \propto \sigma^{4}$. Shortly afterwards, the Kormendy relation \citep{Kor77} connected size (more specifically the half-light or effective radius $R_{e}$) with surface brightness $\mu$. Originally this was $\mu$ \emph{at} $R_{e}$, but now the mean value of $\mu$ within $R_{e}$ is used: $\mu = -2.5\log{(L/R_{e}^2)} + c$.

\citet{DjoDav87} coined the term ``fundamental plane'' (FP for short) to describe the three-dimensional scaling relation between $R$, $\sigma$ and $\mu$ as $R \propto \sigma^a\mu^b$; \citet{DreLynBur87} discovered the same relation independently and concurrently. The FP effectively combines the Faber-Jackson and Kormendy relations as $\mu$ is interchangeable with $L$. In optical passbands, it has since consistently been found that the value of the coefficient $a$ lies in the range 1--1.5, and $b$ from 0.25 -- 0.35. That same year, Faber (1987) pointed out that the virial theorem can be rewritten as $R \propto \sigma^2\mu^{0.4}$ under the assumption of homology - a relation very similar to the FP but with larger values for the coefficients $a$ and $b$. The FP's deviation from the virial scalings implies a scaling of the mass-to-light ratio with FP parameters, now commonly referred to as the ``tilt'' of the fundamental plane.

The tilt and small scatter of the FP are now recognized as strong constraints on models for the formation of elliptical galaxies; however, no firm consensus has yet been reached on the importance of various mechanisms (e.g. hierarchical merging, dissipation, variations in stellar populations) on the FP and its tilt. In this paper, we use numerical simulations to test the hypothesis that ellipticals form from mergers of spiral galaxies in groups, and more specifically that these ellipticals lie on an FP with appropriate tilt and scatter. We begin with a review of the FP and its tilt (\secref{review}), as well as of previous results in this field. \secrefword{methods} summarizes the methods, simulations and observational data used throughout the paper. Measurements of the fundamental plane (\subsecref{fp}) and its tilt (\subsecref{classictilt}) are provided next, along with a full examination of the virial FP (\subsecref{virfp}). We derive dynamical masses from the virial FP in \subsecref{mdyn} and verify the previous results with simple, spherical bulge plus halo models in \subsecref{modelcheck}. In \secref{alttilt}, we introduce an alternative parameterization of the tilt, along with a novel interpretation of its origin (\secref{origin}). \secrefword{discussion} addresses some of the details of the implications of these findings, which are summarized in \secref{conclusions}. The appendix details the sensitivity of the simulated fundamental plane terms to initial conditions (\appref{randics}), specifically the orbital configurations of galaxies within each group.%Two appendices are provided with further data on the run of the virial ratio with radius within sample galaxies (\appref{virratvsr}), as well as the sensitivity of the simulated fundamental plane terms to initial conditions (\appref{randics}), specifically the orbital configurations of galaxies within each group.
\section{Review}
\label{sec:review}
\subsection{The Tilt of the Fundamental Plane}
\label{subsec:tiltreview}
Throughout this paper, we will take the liberty of assuming a consistent set of units of size (kpc), velocity ($\mathrm{km\ s^{-1}}$), mass ($\mathrm{M_{\odot}}$), luminosity ($\mathrm{L_{\odot}}$), and surface brightness ($\mathrm{mag/arcsec^{2}}$), and omit these units from logarithms, writing:
\begin{equation}\label{eqn:fp}
\log{R} = a\log{\sigma} + b\mu + c, \textrm{or:}
\end{equation}\begin{equation}\label{eqn:fpl}
\log{R} = \alpha\log{\sigma} + \beta\log{L} + \gamma, \textrm{where:}
\end{equation}
\begin{equation}\label{eqn:fpcoeff}
\alpha = a/(1-5b),\ \beta = -2.5b/(1-5b),
\end{equation}
and $\gamma$ and $c$ are related by a unit-dependent constant. These relations can be tied to the scalar virial theorem (SVT) for a relaxed object, which states that $2T+W=0$, where $T$ and $W$ are the total kinetic and potential energy, respectively. We distinguish between the SVT and the full (or tensor) virial theorem, which contains additional terms. Now if $T \propto M\sigma^2$, and $W \propto M^2/R$, then $M\sigma^2 \propto M^2/R$, and $R \propto M/\sigma^2$. This yields the ``virial FP'' or virial scaling:
\begin{equation}\label{eqn:fpvi}
\log{R} = -2\log{\sigma} + \log{M} + c_{v}.
\end{equation}
A simple dimensional equality of the form $\sigma^2 = kGM^2/R$ yields the same result, but with $c_{v} - \log(kG)$. The term $k$ is often referred to as the ``structural non-homology'' parameter, because it is constant for self-similar (homologous), virialized systems, but can vary if the assumptions that yielded \eqnref{fpvi} do not hold. It is also referred to as the ``virial parameter'', because its definition is an equality between kinetic and potential energy. \eqnref{fpvi} can also be written as:
\begin{equation}\label{eqn:fpv2}
\begin{split}
\log{R}& = -2\log{\sigma} + \log{L} + \log(M/M_{\star}) \\ 
& + \log(M_{\star}/L) + \log(kG),
\end{split}
\end{equation}
where $M_{\star}/L$ is the stellar mass-to-light ratio (also written as $\Upsilon_{\star}$), whereas $M_{\star}/M$ is the stellar mass fraction within $R$. If these two terms are constant, and if the assumptions of virial equilibrium and the scalings of $T$ and $W$ that yielded \eqnref{fpvi} all hold, then equating \eqnref{fpvi} and \eqnref{fpv2} implies that $\alpha=-2$ and $\beta=1$. These values are the so-called ``virial'' coefficients, which define the ``virial'' FP. If the assumption of homology is broken, then these terms could vary with $L$, $R$ or $\sigma$, and $\alpha$ and $\beta$ will likely differ from -2 and 1, indicating a tilt in the FP.

Up to this point, we have not been specific about the exact definitions of $M$, $L$, $R$ or $\sigma$. To some extent the definitions are arbitrary, especially for the size $R$. We choose to adopt the parametric \cite{Ser68} profile fit for consistency with many observational catalogs, and because it provides a good fit to the surface brightness profiles of elliptical galaxies \citep{CaoCapDOn93,GraCol97}. The Sersic profile defines a total luminosity $L$ for the galaxy, as well as an effective radius $R_{e}$ containing half of the light, and has been shown to be a suitable parameterization for the surface brightness profiles of elliptical galaxies \citep[for a reference to Sersic-related quantities, see][]{GraDri05}. In keeping with convention, the dispersion $\sigma$ is the central, luminosity-weighted, line-of-sight dispersion within $R_{e}/8$, whereas the luminosity-weighted dispersion within $R_{e}$ is written as $\sigma_{e}$.

Observers often prefer to use surface brightnesses rather than luminosities, as surface brightnesses are (nearly) distance-independent and allow for the calibration of the FP as a distance indicator. We use the conventional mean surface brightness within $R_{e}$:
\begin{equation}
\mu_{e} = -2.5\log[(L/2)/(\pi R_{e}^{2})] + 21.572 + 15 + M_{P,\odot},
\end{equation}
where $M_{\odot,P}$ is the absolute magnitude of the sun in a given photometric band $P$. The factor of 15 is for units of $\mathrm{L_{\odot}}$ and kpc. Substituting into \eqnref{fpv2} gives:
\begin{equation}\label{eqn:fp3}\begin{split}
& \log{R_{e}} = 2\log{\sigma} + 0.4\mu_{e} - \log(M_{\star}/L) - \log{k} \\
& + \log(M_{\star}/M) - \log(G) - 0.4M_{\odot,P} - 15.427.
\end{split}
\end{equation}
This is exactly the fundamental plane relation, plus four tilt terms and a fixed constant offset to return it to the virial scalings. The tilt terms are not unique and can be further subdivided, or combined. For example, \cite{HydBer09a} write the non-homology term $k$ as a ratio between dynamical and total mass, $M_{dyn}/M_{tot}$ \citep[see also][]{DOnValSec06}.

Finally, one can use a rotation-corrected velocity $V_{rms}$ \citep[e.g.][]{CapBacBur06} in place of $\sigma$. We define a similar term $S$ as $S^{2} = \sigma_{e}^2 (1 + \langle v/\sigma \rangle^2)$, where $\sigma_{e}$ is the luminosity-weighted dispersion within $R_{e}$ rather than the central dispersion \citep[$S$ is much like the $S_{1}$ of][]{WeiWilFab06}. The mean $v/\sigma$ can be measured in a number of different ways depending on the source of the data; we opt for a similar luminosity-weighted average within $R_{e}$ for consistency with \citet{CapEmsKra11}. Incorporating $S$ adds a tilt term to \eqnref{fpv2}:
\begin{equation}\label{eqn:fpv3}
\begin{split}
\log{R_{e}}= & - 2\log{\sigma} + \log{L} + \log(M_{\star}/L) + \log{k_{S}} \\
       & - \log(M_{\star}/M) + 2\log(\sigma/S) + \log(G),
\end{split}      
\end{equation}
where $k_{S} = S^{2}R/(GM) = k(S/\sigma)^2$.

\subsection{Results from Observations}

Since \citet{DjoDav87}, numerous works have expanded on the interpretation of the FP and its tilt, both from theoretical considerations and observationally attempting to decompose the various contributors to the tilt. \citet{BenBurFab92} proposed an alternative coordinate system dubbed k-space, emphasizing the need to consider both face-on and edge-on projections of the FP, while \citet{GuzLucBow93} quantified the finite extent of the FP. \citet{BenBurFab93} and \citet{GuzLucBow93} were amongst the first to detail the importance of stellar populations (age and metallicity) in determining the $M_{\star}/L$ term's contribution to the tilt. It is now commonly accepted that stellar mass-to-light ratio term is a significant contributor to the tilt, while some form of structural non-homology \citep{PruSim96} can explain the remainder - possibly rotational support, non-universal stellar mass profiles \citep{PruSim97,BusCapCap97,TruBurBel04} and/or variable dark matter fractions \citep{HydBer09b,GraFab10}.

More recently, \emph{total} masses have been inferred by studies using spatially-resolved kinematics and dynamical models \citep[e.g.]{RixdeZCre97,GerKroSag01,CapBacBur06,CapScoAla13} or from strong gravitational lensing \citep[e.g.]{BolTreKoo08,AugTreBol10}. Such analyses have been complicated by the fact that stellar masses are sensitive to the assumed form of the IMF, which recent evidence suggests may be non-universal \citep[e.g.;][]{vDo08,ConvDo12}, impacting the tilt of the FP \citep{DutConvdB11}. Stellar masses can also be sensitive to the star formation history of each galaxy \citep{AllHudSmi09,GraFab10}, which is difficult to constrain. Even if the stellar mass is well-constrained, dynamical and lensing models necessarily make non-trivial assumptions to determine the total mass.

Nonetheless, \citet{AugTreBol10} modelled strong lensing galaxies to find a relation between projected mass density $\Lambda$ within $R_{e}/2$, and the mean $\sigma$ within $R_{e}/2$, $\sigma_{e/2}$, of $\log{R_{e}} \propto (1.83 \pm 0.13)\log{\sigma_{e/2}} - (1.30 \pm 0.06)\log{\Lambda}$. This is equivalent to $\log{R} \propto \alpha\log{\sigma_{e}} + \beta\log{M}$ with $\alpha = -1.14$ and $\beta = 0.81$, neither pair being exactly the virial FP coefficients.

In an independent sample of local early-type galaxies, \cite{CapScoAla13} inferred total masses from dynamical modeling of spatially resolved stellar kinematics. This yielded a best-fit plane of the form $\log{M_{model}} \propto \alpha\log{\sigma_{e}} + \beta\log{R_{e}}$, with $\alpha=1.928 \pm 0.026$ and $\beta=0.991 \pm 0.024$, which is very nearly the virial FP. While this would seem to imply that the FP is simply a reflection of the fact that galaxies are virialized, recall that \eqnref{fpvi} required the assumptions that $M\sigma^{2} \propto T$, and that $M^{2}/R \propto T$, neither of which have been validated. Indeed, \cite{CapScoAla13} point out that the fact that $k$ appears to be constant for their particular mass and radius definitions is not necessarily an obvious extension of the virial theorem, since real galaxies can be complex, multi-component systems. We will return to this point later.

\subsection{Results from Theory}

Galaxy mergers have long been thought to form elliptical galaxies, and numerical simulations - especially $N$-body simulations - have been a particularly powerful tool for testing this hypothesis (see \citet{Str99} and \citet{Str06} for excellent reviews). However, a much smaller fraction of the literature has attempted to predict or explain the nature of the FP and its tilt. \citet{WeiHer96} presented detailed simulations of mergers in groups, but their predictions for the FP were left unpublished \citep[see][]{Wei95}. \citet{Bek98} presented one of the first analyses of the FP tilt in dissipational binary spiral merger simulations, suggesting that dissipation could establish non-homology. A number of studies alternatively focused on mergers of spheroidal galaxies, including \citet{CapdeCCar95}, \citet{NipLonCio03}, \citet{BoyMaQua05,BoyMaQua06} and more recently \citet{HilNaaOst12}. Most of these studies have focused on determining whether the FP and its tilt can be preserved in subsequent mergers of spheroids - so far, the answer is ``maybe,'' depending on the number of mergers, mass ratios, orbital parameters and properties of the merging systems. However, such simulations typically do not address how spheroidal galaxies are formed in the first place.

Binary, major mergers of spiral galaxies are still viewed as a plausible formation channel for elliptical galaxies, although this mechanism is not universally accepted \citep{NaaOst09}. \citet{AceVel05} presented simulations of binary mergers of spirals scaled to follow the observed Tully-Fisher relation \citep{TulFis77}. These merger remnants were similar to ellipticals and followed a tilted FP even in the absence of gas dissipation, owing to the precise scaling of the progenitor spirals. By contrast, \citet{RobCoxHer06} reported that dissipational binary mergers were sufficient to create a tilted FP ($a=1.55$ and $b=0.33$) with 0.06 dex scatter. \citet{HopCoxHer08} went a step further in claiming that dissipation is both necessary and sufficient to create the FP. However, such binary merger simulations did not incorporate truly hierarchical cosmological merger histories and so did not consider the impact that repeated, mainly minor mergers may have on the size and mass growth of ellipticals. Although a few consecutive re-mergers of merger remnants were included, binary mergers alone do not capture the full range of possible merger histories of ellipticals, and so are at risk of overestimating the role of dissipation over structural non-homology from dissipationless merging, as reported by \citet{AceVel05}.

A relatively new class of cosmological ``zoom'' simulations can in principle follow realistic merger histories and form elliptical galaxies self-consistently. Zoom simulations have been used to study 2D scaling relations \citep{NaaJohOst09,OseNaaOst12} and detailed kinematics \citep{NaaOseEms14}, but not the fundamental plane, as far as we are aware - perhaps due to the necessarily limited resolution of full cosmological simulations. \citet{OnoDomSai06} presented scaling relations from hydrodynamical cosmological simulations in small volumes (10 Mpc-side boxes), but with low resolution by today's standards ($64^3$ gas and dark matter particles each). \citet{FelCarMay11} and \citet{DubGavPei13} compared their central group galaxies to the projected fundamental plane, but with 6--7 galaxies each, neither study had the statistics to fit scaling relations. Hybrid simulations are an alternative to full ab initio conditions, whereby halos in a dark-matter only simulation are seeded with realistic galaxies at intermediate redshifts and evolved to the present \citep[e.g.][]{Dub98}. \citet{NipStiCio03} simulated the evolution of 5 massive spheroidal galaxies in a $z=0.59$ cluster, while \citet{RusSpr09} analyzed several dozen merger dry merger remnants in a rich cluster seeded with 50 galaxies; in both cases, the initial galaxies contained a stellar bulge and dark halo but not disks. Both papers presented plots of the projected FP, finding evidence for a tilt - in \citet{RusSpr09}, driven by variable dark matter fractions - but neither quoted best-fit FP parameters.

\citet[][hereafter Paper I]{TarDubYee13} presented controlled, collisionless N-body simulations of mergers of spirals in groups, incorporating cosmologically-motivated merger histories and measuring accurate sizes, dispersions and luminosities for the central merger remnants. We now aim to extend the analysis of the morphologies and kinematics of these galaxies to assess the impact of dissipationless merging of spiral galaxies on the FP.

\section{Methods}
\label{sec:methods}

This work combines the results of simulations and observations, using similar methods for each data set. All sizes, ellipticities and Sersic indices are derived via two-dimensional, single Sersic profile fits, where possible. Unless otherwise specified, we define $R_{e}$ as $\sqrt{ab}$, where $a$ and $b$ are the major and minor axes of the best-fit Sersic ellipse, respectively. In this way, $R_{e}^2$ is proportional to the area containing half of the projected Sersic model luminosity. Similarly, all projected quantities measured within $R_{e}$ are measured within the best-fit ellipse, not a circle. By convention, the central dispersion $\sigma$ is defined as the projected velocity dispersion within $R_{e}/8$; we refer to the luminosity-weighted dispersion within $R_{e}$ as $\sigma_{e}$. All magnitudes and luminosities are for the SDSS $r$-band.

We refer to the projected total mass within the ellipse defined by $R_{e}$ and the ellipticity $e$ as $M_{R_{e},2D}$ - similar to a lensing mass. We also define the mass enclosed within a sphere of radius $\langle R_{e} \rangle$ as $M_{R_{e,3D}}$, where $\langle R_{e} \rangle = \sqrt[3]{R_{maj}R_{med}R_{min}}$ - the cube root of the product of $R_{e}$ from each principal axis projection. $M_{R_{e,3D}}$ is then the mass contained within a sphere which encloses roughly half of the galaxy's light in any given projection, and slightly less than half in three dimensions.

For fitting of all relations including the fundamental plane, we perform linear least-squares regression using MATLAB's svd (singular value decomposition) function. Errors on the fit parameters are estimated from bootstrapping with random resampling, allowing for duplicates and using at least 1,000 bootstrapped samples. As in Paper I, we also perform weighted fits to compensate for the shallower luminosity function of the simulated galaxies, using MATLAB's svds function.

\subsection{Simulation Data}
\label{subsec:simulations}

The simulations of Paper I were run with the PARTREE N-body tree code \citep{Dub96}. Briefly, they consist of over a hundred galaxy groups, initially comprising three to thirty spiral galaxies sampled from a realistic luminosity function. Each group was designed to collapse like a high redshift (z=2) group at the turnaround radius, inducing mergers and forming a central elliptical. Unlike previous merger simulations, the groups were meant to sample a variety of plausible cosmological accretion histories, albeit not an unbiased sample. The initial galaxies were also chosen to follow the observed Tully-Fisher relation \citep{TulFis77} with no intrinsic scatter.

The initial conditions in the simulations were tightly controlled, allowing only a few parameters to vary. All of the galaxies were self-similar, re-scaled models of M31 \citep{WidPymDub08}. Two versions of each simulation were run: one where the spirals began with an exponential bulge (the ``bulge Sersic index=1'' or $B.n_{s}=1$ sample), and another otherwise identical run with de Vaucouleurs profile bulges ($B.n_{s}=4$). This roughly brackets the range of realistic bulge profile in spiral galaxies, allowing us to test the impact of input galaxy structure on the properties of the final remnant. Additionally, groups of a given mass were seeded with varying numbers of galaxies, in order to test the impacts of multiple, mostly minor mergers versus a few mostly major mergers. We distinguish between these subsamples as ``Many'' (M) and ``Few'' (F), where the former groups began with a larger-than-average number of galaxies for their mass. Similarly, the main group sample had galaxy luminosities drawn from a realistic luminosity function (``LF''), whereas a smaller control sample featured equal-mass mergers (``Eq'').

Although collisionless, the sample size and resolutions (typically 1--2 million stellar particles with a 100 pc softening radius) of the Paper I simulations are both higher than in typical zoom simulations \citep[e.g.]{NaaJohOst09}. Each central remnant was imaged along ten randomly oriented, evenly-spaced projections, as well as the three principal axis vectors, creating surface brightness and kinematic maps. A full account of the morphology, kinematics, and structural properties of the simulated galaxies in this catalog is available in Paper I, along with a more detailed description of the methodology.

\subsection{Observational Data}
\label{subsec:observations}

We use 2D Sersic fits from \citet[][hereafter S+11]{SimMenPat11} and visual morphological classifications from \citet[][hereafter N+10]{NaiAbr10}, both based on SDSS data. We also use spatially resolved kinematics for 65 nearby ellipticals from ATLAS3D \citep{CapEmsKra11}, including kinematic measures from \citet{CapScoAla13}, and stellar mass-to-light ratios and dark matter fractions from \citet{CapMcDAla13}. In Paper I, we used 2D GALFIT fits described in the appendix of \citet{KraAlaBli12}. In this paper, we will use the values of $L$ and $R_{e}$ tabulated in \citet{CapScoAla13}, as recommended by the authors of that paper. Although these are not derived from 2D Sersic fits, they are generally similar for most galaxies and allow for more direct comparison to \citet{CapScoAla13}. Since spatially resolved kinematics are unavailable for most SDSS galaxies, we use central velocity dispersions $\sigma$ unless otherwise specified.

Wherever possible, we have opted to use similar methodologies in all cases. The simulation sizes and luminosities are based on single Sersic fits using GALFIT \citep{PenHoImp02,PenHoImp10}. SDSS data use single Sersic fits from GIM2D \citep{SimWilVog02}. SDSS stellar masses are based on fits to photometry \citep{MenSimPal14}, using a \citet{Cha03} initial mass function (IMF). These stellar masses are on average marginally but not significantly different from those used in Paper I, and we refer to \citet{MenSimPal14} for full details. A3D stellar masses are derived by multiplying the model $L_{r}$ with the $M/L_{r}$ derived from fits to spatially resolved spectra assuming a \cite{Sal55} IMF \citep{CapMcDAla12,CapMcDAla13}. To maintain consistency with the normalization of $M/L$ values with different IMFs, we divide the A3D $M/L_{r}$ by a factor of 1.7, which compensates for the systematically larger $M/L_{r}$ in a Salpeter IMF for an old, solar-metallicity population. This does not entirely account for systematic differences between stellar masses from the two samples, and \citet{MenSimPal14} caution that stellar masses derived from photometric fits are subject to up to 60\% systematic errors. We similarly caution against over-interpretation of trends based on stellar masses with a fixed IMF.

\section{Results}
\label{sec:results}

\subsection{The Fundamental Plane}
\label{subsec:fp}

The FP relation is typically written as $\log R_{e} = a\log{\sigma} + b\mu + c$, as in \eqnref{fp}. We derived this relation from the SVT in the traditional way in \subsecref{tiltreview}. The FP can also be expressed in terms of $L$ instead of $\mu$. This formulation is equivalent to using $\mu$, and \eqnref{fpcoeff} gives simple transformations between the luminosity and $\mu$ FPs. While luminosity is arguable a more fundamental variable, being independent of $R_{e}$, we will use $\mu$ for the moment for consistency with previous studies.

\begin{figure*}
\includegraphics[width=0.49\textwidth]{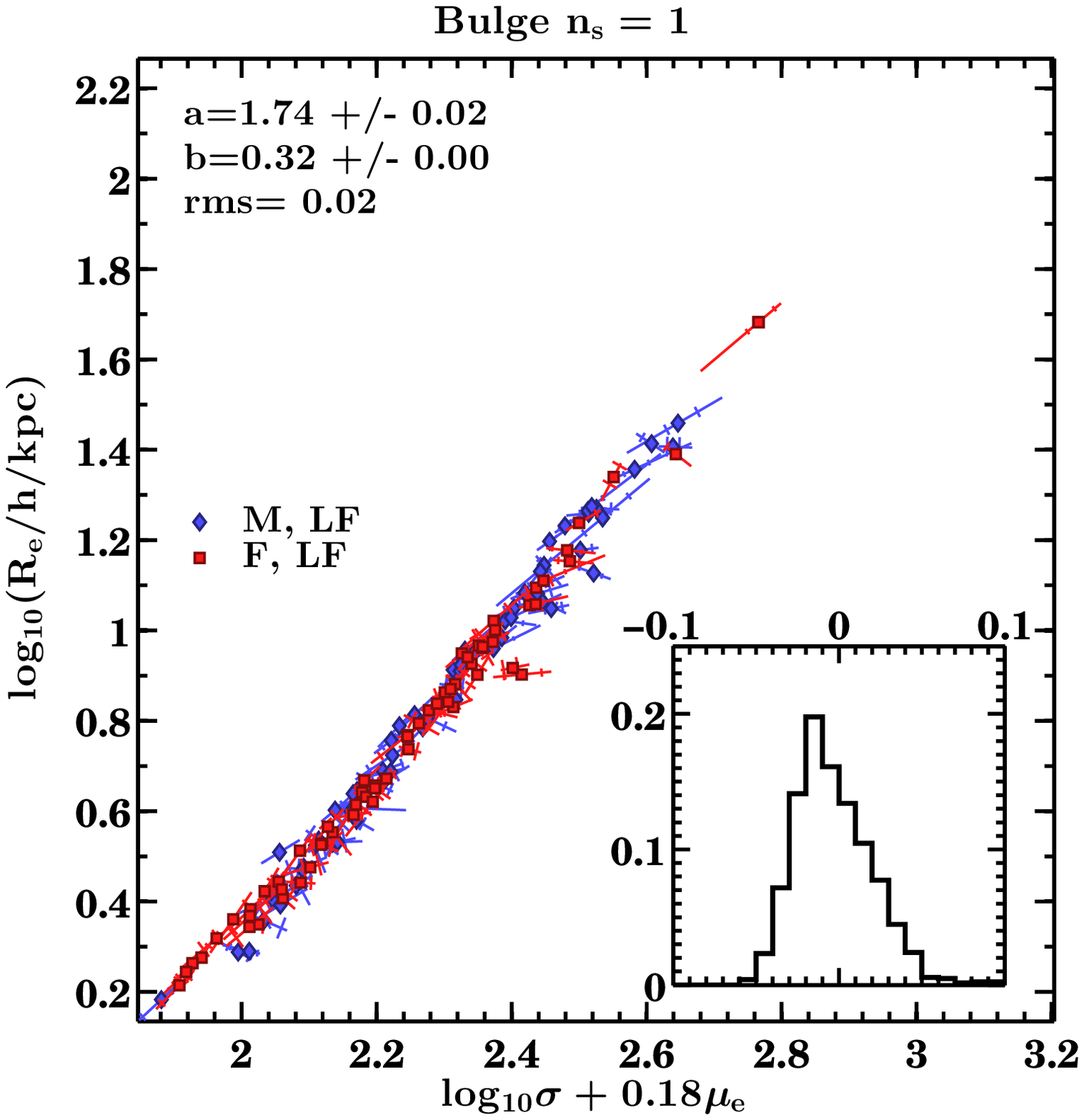}
\includegraphics[width=0.49\textwidth]{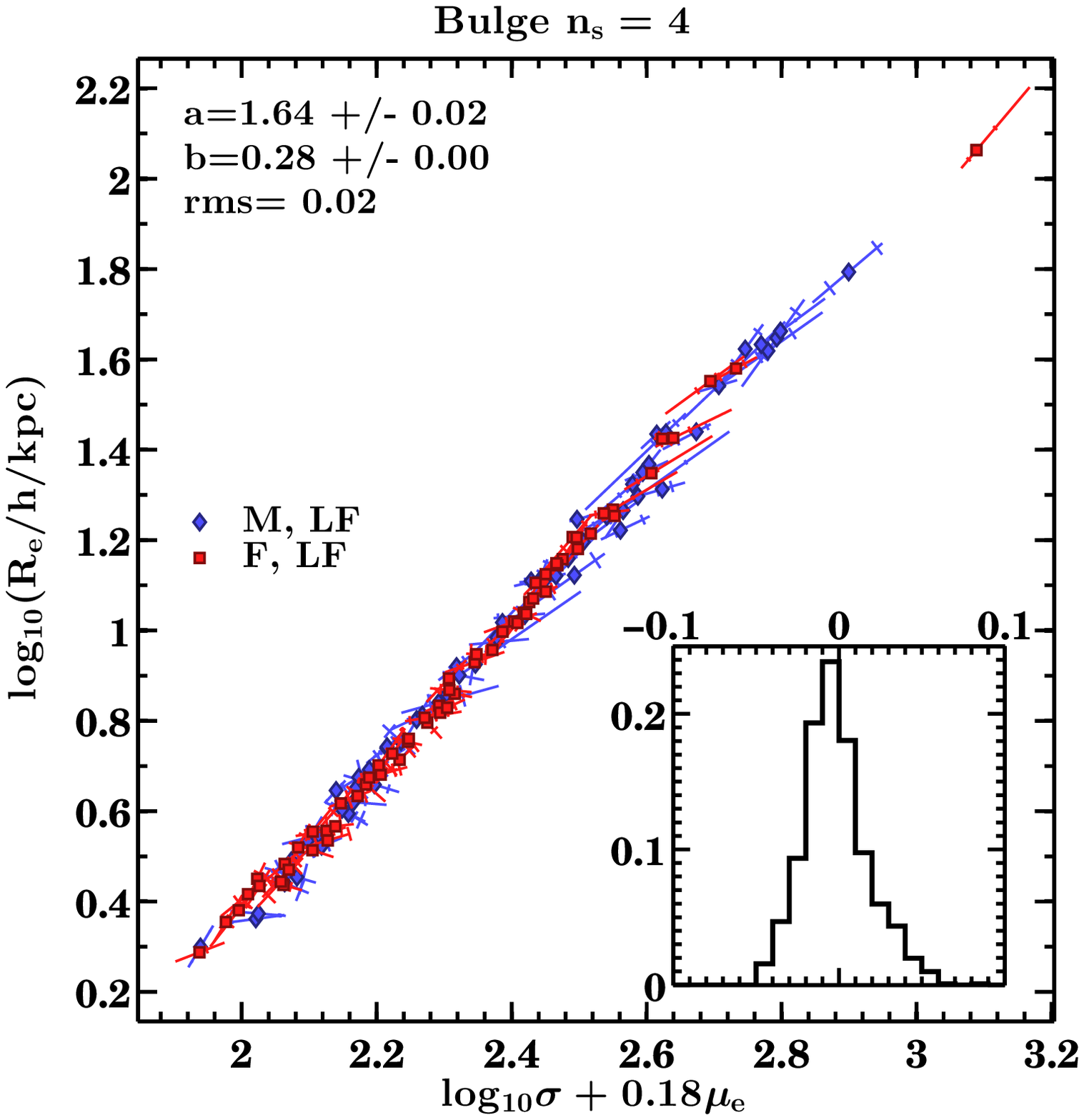}
\caption{Projected FP for the B.$\mathrm{n_{s}}=4$ and 1 samples, plotted in the style of \citet{HydBer09a}, where the $\mu_{e}$ term in the x-axis includes only deviations from $\langle \mu_{e} \rangle$ (i.e. it is $\mu_{e} - \langle \mu_{e} \rangle$). The coefficient 0.18 matches the average value of $b/a$ for both panels. Different projections of the same group are shown as lines of best fit, with the data point marking the median projection. The length of the line shows the range of values from 10 random projections. Perpendicular lines cross at the 25th and 75th percentiles, with a length equivalent to the rms dispersion of points perpendicular to the line of best fit. Inset shows the PDF of orthogonal scatter from the best-fit FP.
\label{fig:fplane_proj}}
\end{figure*}

\figref{fplane_proj} shows the projected FP, with two variables ($\sigma$ and $\mu_{e}$) collapsed onto the x-axis, as in \cite{HydBer09a}. This visually demonstrates the exceptionally small orthogonal scatter of the FP, even for the exponential bulge ($B.n_{s}=1$) sample, which is not shown exactly edge-on. Similarly, the tilt does depend on the structure of the progenitors, since it is larger in magnitude for the $B.n_{s}=4$ sample, as is the extent of the FP.

\begin{table}
\centering
\caption{Sersic model Fundamental Plane fits}
Simulations: Ten equally-spaced projections, randomly oriented \\
\begin{tabular}{cccccc}
\hline 
B.$\mathrm{n_{s}}$ & Subsample & $a$ & $b$ & Intercept & R.M.S. \\
\hline
1 & All & 1.74 & 0.32 & -9.50 $\pm$ 0.06 & 0.02 \\
%1 & Weighted & 1.74 & 0.32 & -9.46 $\pm$ 0.07 & 0.02 \\
1 & Many & 1.80 & 0.31 & -9.49 $\pm$ 0.10 & 0.02 \\
1 & Few & 1.70 & 0.31 & -9.26 $\pm$ 0.11 & 0.02 \\
\hline
4 & All & 1.64 & 0.29 & -8.65 $\pm$ 0.04 & 0.02 \\
%4 & Weighted & 1.64 & 0.28 & -8.62 $\pm$ 0.04 & 0.02 \\
4 & Many & 1.67 & 0.28 & -8.60 $\pm$ 0.06 & 0.02 \\
4 & Few & 1.64 & 0.29 & -8.62 $\pm$ 0.08 & 0.02 \\
\hline
All & All & 1.69 & 0.29 & -8.89 $\pm$ 0.04 & 0.02 \\
%All & Weighted & 1.69 & 0.29 & -8.83 $\pm$ 0.04 & 0.02 \\
All & Many & 1.75 & 0.28 & -8.71 $\pm$ 0.05 & 0.02 \\
All & Few & 1.66 & 0.29 & -8.85 $\pm$ 0.07 & 0.02 \\
\hline
\end{tabular}

Principal axis projections, unweighted \\
\begin{tabular}{cccccc}
\hline 
B.$\mathrm{n_{s}}$ & Projection & $a$ & $b$ & Intercept & R.M.S. \\
\hline 
1 & Major axis & 1.76 $\pm$ 0.03 & 0.30 & -9.22 $\pm$ 0.25 & 0.03 \\
1 & Minor axis & 1.71 $\pm$ 0.02 & 0.31 & -9.30 $\pm$ 0.29 & 0.02 \\
4 & Major axis & 1.71 $\pm$ 0.02 & 0.28 & -8.74 $\pm$ 0.14 & 0.02 \\
4 & Minor axis & 1.58 $\pm$ 0.01 & 0.28 & -8.36 $\pm$ 0.12 & 0.01 \\
\hline
\end{tabular}

Observations \\
\begin{tabular}{ccccccc}
\hline 
Cat. & T. & W. & $a$ & $b$ & Intercept & R.M.S. \\
\hline 
S+11 & E & N & 1.35 $\pm$ 0.00 & 0.28 & -7.89 $\pm$ 0.01 & 0.06 \\
S+11 & E & Y & 1.28 $\pm$ 0.00 & 0.29 & -7.95 $\pm$ 0.01 & 0.06 \\
\hline
N+10 & E & N & 1.36 $\pm$ 0.04 & 0.22 & -7.88 $\pm$ 0.12 & 0.06 \\
N+10 & S0 & N & 1.08 $\pm$ 0.02 & 0.27 & -7.11 $\pm$ 0.08 & 0.07 \\
N+10 & E & Y & 1.08 $\pm$ 0.10 & 0.28 & -7.40 $\pm$ 0.28 & 0.06 \\
%N+10 & S0 & Y & 0.87 $\pm$ 0.02 & 0.26 & -6.63 $\pm$ 0.12 & 0.08 \\
\hline
A3D & E & N & 1.29 $\pm$ 0.09 & 0.33 & -9.12 $\pm$ 0.55 & 0.07 \\
A3D & S0 & N & 0.96 $\pm$ 0.07 & 0.30 & -7.63 $\pm$ 0.38 & 0.07 \\
A3D & E & Y & 1.31 $\pm$ 0.09 & 0.33 & -9.09 $\pm$ 0.52 & 0.07 \\
%A3D & S0 & Y & 0.98 $\pm$ 0.09 & 0.31 $\pm$ 0.02 & -7.89 $\pm$ 0.49 & 0.09 \\
\hline
\end{tabular}
\tablecomments{Slopes are given in log space, i.e., for $\log R_{e}$ as a function of $\log L$. Simulation data are from analyses after 10.3 Gyr, including various subsamples of randomly oriented (but equally spaced) projections, as detailed in the text, as well as principal axis projections. Simulated data include subsamples for relatively many or few mergers. Observational data for each catalog (Cat.) and Hubble type (T.) are 1/V$\mathrm{_{max}}$ corrected, with fits optionally weighted (W.) or not by the difference between the simulated and observed luminosity functions. R.M.S. lists the r.m.s. orthogonal scatter of all points from the best-fit relation. Errors on $a$ for simulations and on $b$ for simulations, N+10 and S+11 are uniformly 0.01 or smaller and are omitted for brevity; for A3D, errors on $b$ are 0.02.}
\label{tab:fpfits}
\end{table}

\tabref{fpfits} tabulates best-fit FP parameters for simulations and observations alike. Several trends are clear from the simulations. First, the tilt is smaller than observed but definitely in the correct sense: $a<2$ and $b<0.4$. The scatter in the simulations is exceptionally small (0.02 dex), even when combining different progenitor subsamples, and is considerably smaller than for the observations (0.06 -- 0.08 dex). For the most part, the different observational samples are consistent with one another, other than unusually small $a$ parameters for the N+10 weighted subsample and $b$ for the unweighted (both possibly due to the undersampling of faint and small ellipticals in N+10). Nonetheless, S+11 is largely consistent with the completely independent measurements from A3D, although the A3D intercepts are slightly lower and scatter slightly larger. It is also clear that the FP of S0 galaxies is quite different from that of ellipticals; since the simulated galaxies show no signs of being genuine bulge plus disk systems, we will not consider further comparisons with S0s. It should be noted, however, that the S+11 sample is likely contaminated with S0s, which is unavoidable since no automated classification can separate them (see Paper I for details).

In Paper I, we presented optional weighting schemes which can have a significant impact on 2D scaling relations - i.e., projections of the FP. For the simulations, we weighted the ``Many'' merger sample more heavily at the more luminous end, and the ``Few'' merger sample at the faint end. This weighting does not impact the simulated FP fits, because all of these points really do lie on a single plane with minimal scatter. No weighting scheme will make a significant difference unless it changes the weights where there is curvature in a scaling relation, and there is no curvature in the simulated FP. We omit these weighted fits from \tabref{fpfits}, as all of the parameters are within the errors of the unweighted fits.

A second weighting scheme in Paper I weighted the \emph{observed} data to match the much flatter LF of the simulations. This scheme can make a significant difference (especially for N+10), which is possible if there is some curvature in the observed plane at the high mass end. In the N+10 case specifically, it may also be due to the sample selection, which selected roughly a log-normal distribution around L* rather than a magnitude- or volume-limited sample.

The FP parameters also depend on the projection angle. Choosing only minor axis projections yields the steepest FP, whereas major axis projections minimize the tilt. They can also be sensitive to the orbits of the merging galaxies - we leave this analysis to \appref{randics}. Regardless, the tilt in the simulations is weaker than observed. This difference can be quantified in various ways. One methods is to project the vector between the unit normal of the simulated plane (-1, $a \approx 1.69$, $b \approx 0.29$) and the unit normal of the virial plane (-1, $a=2$, $b=0.4$) onto a similar vector between the observed (-1, $a \approx 1.3$, $b \approx 0.29$) and virial plane's unit normals. This gives a ``tilt fraction'' of 37\%. Any other quantification of the tilt fraction would yield a similar result, and so this difference requires an explanation.

\begin{table}
\centering
\caption{Sersic model stellar mass Fundamental Plane fits}
Simulations: Ten equally-spaced projections, randomly oriented \\
\begin{tabular}{cccccc}
\hline 
B.$\mathrm{n_{s}}$ & Subsample & $a$ & $b$ & Intercept & R.M.S. \\
\hline
All & All & 1.72 & 0.31 & -9.07 $\pm$ 0.04 & 0.02 \\
%All & Weighted & 1.72 & 0.31 & -9.00 $\pm$ 0.03 & 0.02 \\
All & Many & 1.78 & 0.30 & -8.91 $\pm$ 0.05 & 0.02 \\
All & Few & 1.69 & 0.32 & -9.05 $\pm$ 0.07 & 0.02 \\
All & All-S & 1.78 & 0.29 & -8.69 $\pm$ 0.05 & 0.02 \\
\hline
\end{tabular}

Observations \\
\begin{tabular}{ccccccc}
\hline 
Cat. & T. & W. & $a$ & $b$ & Intercept & R.M.S. \\
\hline 
S+11 & E & N & 1.48 $\pm$ 0.00 & 0.28 & -7.87 $\pm$ 0.01 & 0.06 \\
S+11 & E & Y & 1.42 $\pm$ 0.01 & 0.28 & -7.96 $\pm$ 0.01 & 0.06 \\
%S+11-OLD & E & N & 1.52 $\pm$ 0.00 & 0.28 & -8.05 $\pm$ 0.01 & 0.06 \\
%S+11-OLD & E & Y & 1.43 $\pm$ 0.01 & 0.29 & -8.03 $\pm$ 0.01 & 0.06 \\
\hline
N+10 & E & N & 1.56 $\pm$ 0.04 & 0.27 & -8.09 $\pm$ 0.11 & 0.06 \\
N+10 & S0 & N & 1.26 $\pm$ 0.02 & 0.28 & -7.76 $\pm$ 0.06 & 0.06 \\
N+10 & E & Y & 1.25 $\pm$ 0.06 & 0.28 & -7.67 $\pm$ 0.20 & 0.06 \\
%N+10 & S0 & Y & 1.17 $\pm$ 0.03 & 0.28 & -7.34 $\pm$ 0.10 & 0.06 \\
\hline
A3D & E & N & 1.71 $\pm$ 0.08 & 0.36 & -10.13 $\pm$ 0.49 & 0.05 \\
A3D & S0 & N & 1.56 $\pm$ 0.09 & 0.28 & -8.29 $\pm$ 0.45 & 0.06 \\
A3D & E & Y & 1.73 $\pm$ 0.08 & 0.36 & -10.10 $\pm$ 0.48 & 0.06 \\
\hline
A3D-S & E & N & 1.78 $\pm$ 0.11 & 0.37 & -10.04 $\pm$ 0.67 & 0.06 \\
A3D-S & S0 & N & 1.87 $\pm$ 0.14 & 0.30 & -9.42 $\pm$ 0.61 & 0.07 \\
A3D-S & E & Y & 1.83 $\pm$ 0.11 & 0.37 & -10.50 $\pm$ 0.61 & 0.06 \\
\hline
\end{tabular}
\tablecomments{Column headings and notes are as in \tabref{fpfits}, but surface brightnesses/magnitudes are now based on stellar masses instead of luminosities. See text for details on sources of stellar masses. A3D-S and All-S use the kinematic measure $S$ in place of $\sigma$, correcting for ordered rotation - see \eqnref{fpv3}.}
\label{tab:fpfitsmstellar}
\end{table}

The discrepancy between the predicted and observed tilt is mainly in the $a$ parameter and is lessened if one considers the stellar mass FP (\tabref{fpfitsmstellar}). This is justifiable because the simulations effectively do not include any stellar population variations and have a constant $M_{\star}/L$, whereas luminous ellipticals tend to have larger $M_{\star}/L$ in general (and $M_{\star}/L_{r}$ in particular), assuming a universal IMF. This implies that a substantial portion of the tilt is due to variations in stellar populations along the FP. Once this is accounted for, the gap between the observed tilt and the $B.n_{s}=4$ sample's tilt is considerably smaller. In fact, for A3D, the observed tilt in the stellar mass FP is \emph{smaller} than in the $B.n_{s}=4$ sample, as A3D has a significantly larger value of $b \approx 0.36$ than the other samples. This $b$ value is consistent with the near-infrared J-band FP fits from the 6dF survey \citep{MagSprCol12} of $a=1.52 \pm 0.03$ and (effectively) $b=0.36 \pm 0.02$. This is close to a stellar mass FP, since $M/L_{\star,J}$ is less sensitive to age and metallicity than $M/L_{\star,r}$.

It is clear that multiple dissipationless mergers can produce a tilt in the FP, and that this tilt could be a significant fraction of the observed tilt. However, we caution again that the stellar mass FP is sensitive to the assumption of a universal IMF and/or star formation history. Furthermore, the stellar masses for A3D are based on fits to spatially resolved spectra assuming a Salpeter IMF, whereas SDSS stellar masses are based on fits to photometry assuming a Chabrier IMF instead; both methods have their caveats and we make no judgment as to which are more likely to be accurate.

Finally, \tabref{fpfitsmstellar} includes a fit to the FP using $S$ in place of $\sigma$, for A3D and the simulations. Using $S$ lowers the tilt slightly in A3D, although only significantly for the S0s - not surprisingly, since they show more rotational support and are more affected by the $v/\sigma$ correction. Using $S$ in the simulations makes little difference in $a$ and actually lowers $b$ to 0.29. This is likely because the simulated ellipticals generally have less rotational support than A3D ellipticals, as shown in Paper I.

\subsection{The Tilt of the Fundamental Plane}
\label{subsec:classictilt}

Having established that the FP of the simulations is tilted from the virial relation, we will now decompose the FP and fit the various tilt terms from \eqnref{fpv2}. We use $M_{r,\odot}=4.68$ throughout, which sets the constant in \eqnref{fpv2} to -11.93.

\begin{figure}
\includegraphics[width=0.49\textwidth]{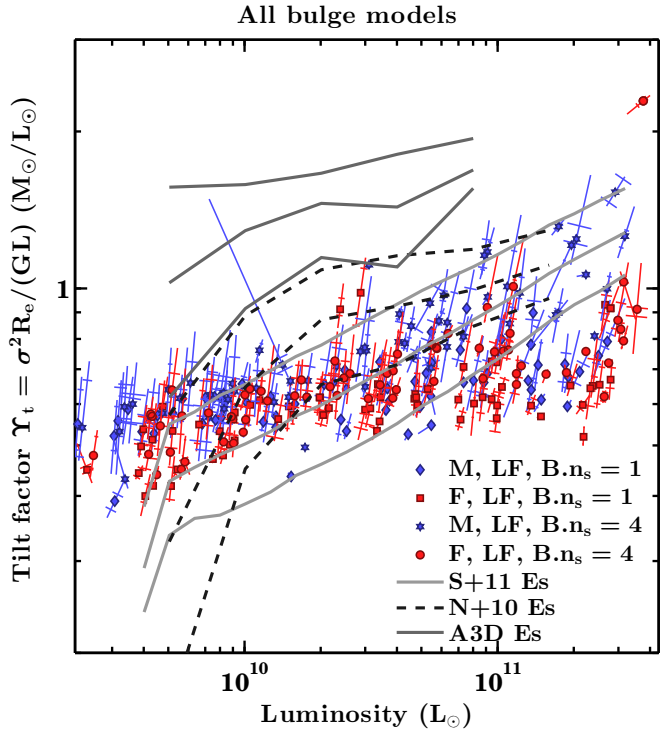}
\caption{The tilt factor $\Upsilon_{t}$ for simulated and observed ellipticals. The lines for observational catalogs separate quartiles. ATLAS3D ellipticals are somewhat discrepant from both the simulated trend and observed trends for SDSS ellipticals, while the visually-classified N+10 ellipticals show a sharp drop at low luminosities (in part because N+10's sample was biased to select brighter galaxies). Nonetheless, all relations show a positive trend.
\label{fig:virialk3d}}
\end{figure}

The full tilt of the FP can be characterized from the equality $\sigma^2 = \Upsilon_{t} GL/R_{e}$, where the tilt factor $\Upsilon_{t}=k(M/L)$ is a mass-to-light ratio that encompasses the entirety of the tilt, including non-homology ($k$), stellar population variations ($\Upsilon_{\star}$) and variable dark matter fractions ($M_{\star}/M_{R_{e,3D}}$). \figref{virialk3d} shows $\Upsilon_{t}$ as a function of luminosity for the entire sample. The term clearly varies with luminosity in both observations and simulations, although the variation is shallower for simulations (unsurprisingly, since the tilt is also smaller). Nonetheless, the intercept and median values of $\Upsilon_{t}$ are consistent with S+11, though somewhat offset from A3D. This is mainly because the simulated galaxies have lower $\sigma$ than observed, and A3D galaxies have slightly larger $\sigma$ than the other samples. Sample selection is evidently important, since the N+10 is a subset of the S+11 samples and uses the same $\sigma$, $R_{e}$ and $L$ measures, but shows an anomalous break in $\Upsilon_{t}$ at $2 \times 10^{11} L_{\star}$. This break is likely to be the cause of the difference between weighted and unweighted N+10 fits. We do not speculate further on the causes of these systematic differences and simply re-assert that the simulated tilt is at least broadly consistent with the observed intercept and shallower in the slope.

\begin{figure*}
\includegraphics[width=0.49\textwidth]{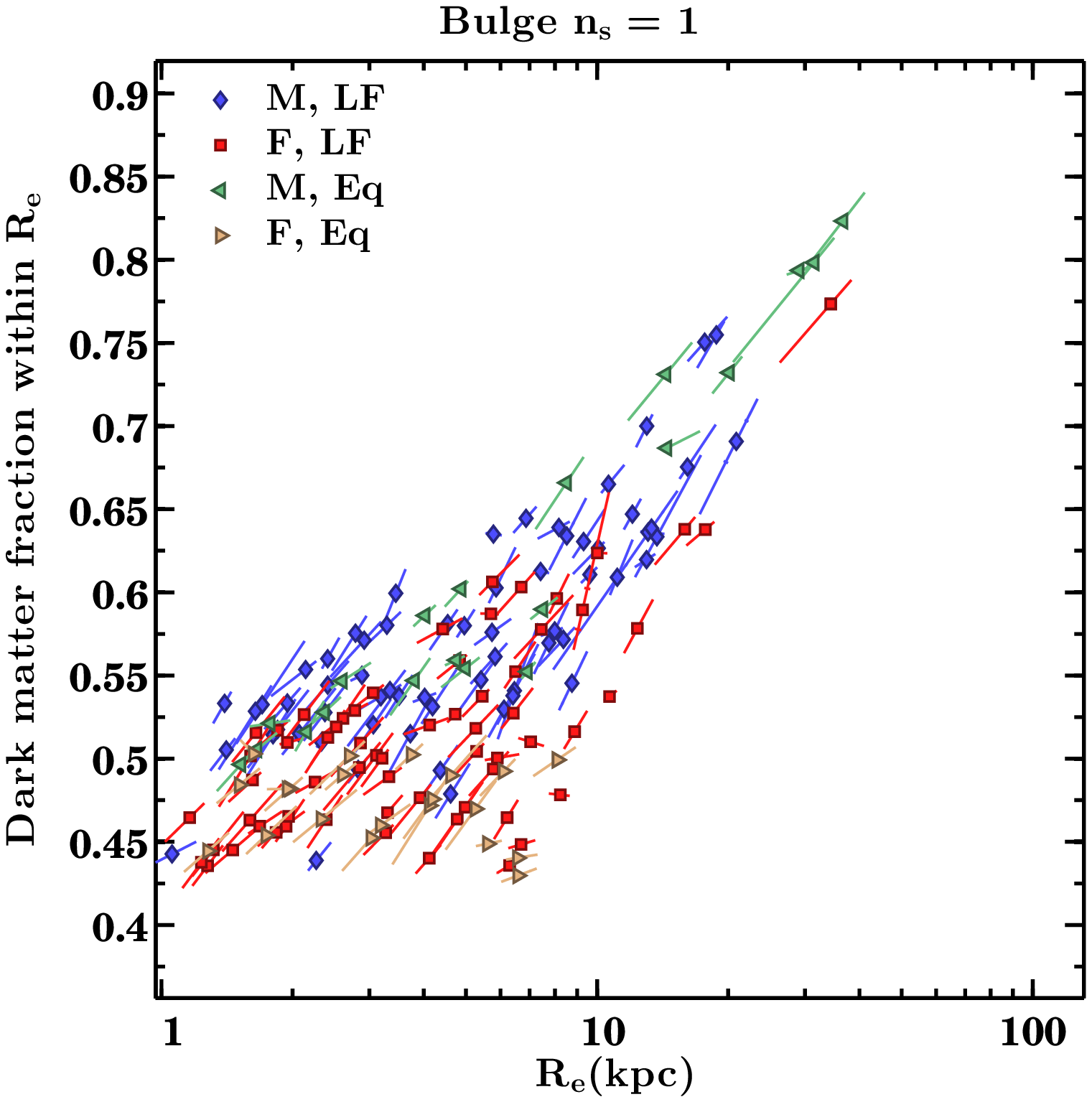}
\includegraphics[width=0.49\textwidth]{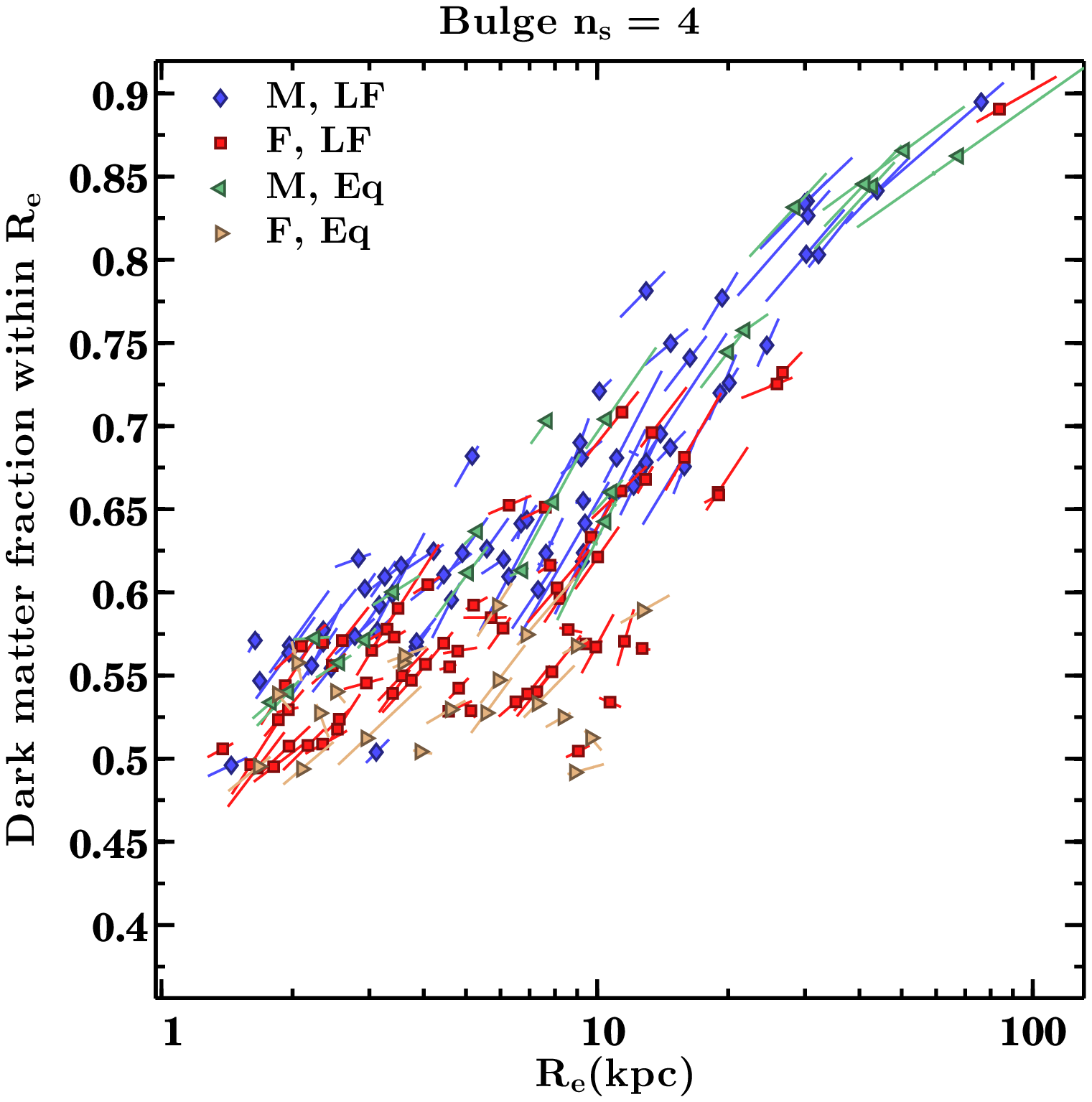}
\caption{Projected dark matter fraction within $R_{e}$. The dependence of the baryon fraction within $R_{e}$ ($M_{\star,R_{e},2D}/M_{R_{e},2D}$ = 1-$M_{DM,R_{e},2D}/M_{R_{e},2D}$) on size and mass is a major contributor to the tilt of the fundamental plane.
\label{fig:fracdm_proj}}
\end{figure*}

We now turn to identifying which of the tilt terms contribute to the variable $\Upsilon_{t}$. \figref{fracdm_proj} shows projected dark matter fractions, i.e. $1-M_{\star,R_{e},2D}/M_{R_{e},2D}$. Similar fractions of order 2--5\% smaller are found for $1-M_{\star,R_{e},3D}/M_{R_{e},3D}$. Dark matter fractions increase for larger galaxies, because the effective radii extend further out to regions dominated by the halo. This is evidenced by the fact that independent projections of the same galaxy with larger $R_{e}$ have larger dark matter fractions - unsurprisingly, since the dark matter fraction will increase with radius as long as the stellar density profile is steeper than the halo profile.

The non-trivial discovery here is that $M_{\star}/M_{R_{e},2D}$ is significantly smaller in larger/more luminous galaxies, driving what would appear to be a large fraction of the tilt. This is partly caused by the fact that, as shown in Paper I, galaxies undergoing many mergers (M) are larger at fixed luminosity than those formed from few mergers (F); similarly, at a fixed size, the M subsample is less luminous and so contains a smaller stellar mass fraction. For the whole sample, we find that $M_{DM,R_{e},3D} \propto R_{e}^2$ nearly exactly, whereas $M_{\star,R_{e},3D} \propto R_{e}^{1.72}$, with a slightly shallower slope for the most luminous galaxies. The enclosed dark matter mass in different galaxies scales more steeply with $R_{e}$ than does the stellar mass.

The full explanation for the origin of this trend in dark matter fractions is complicated by the fact that neither the stellar density profiles nor the dark halo profiles are self-similar in the merger remnants. Nonetheless, we have demonstrated that merging multiple self-similar galaxies produces a particular scaling relation for $M_{\star}/M$ within $R_{e}$, rather than retaining the same constant fraction that they began with (roughly 30--35\%, depending on how one defines $R_{e}$ for a spiral galaxy). We will discuss the absolute values of these dark matter fractions and compare them to observations further in \secref{discussion}; for now, what matters is that the trend lies in the correct sense to cause some part of the tilt in the FP.

\begin{figure*}
\includegraphics[width=0.34\textwidth]{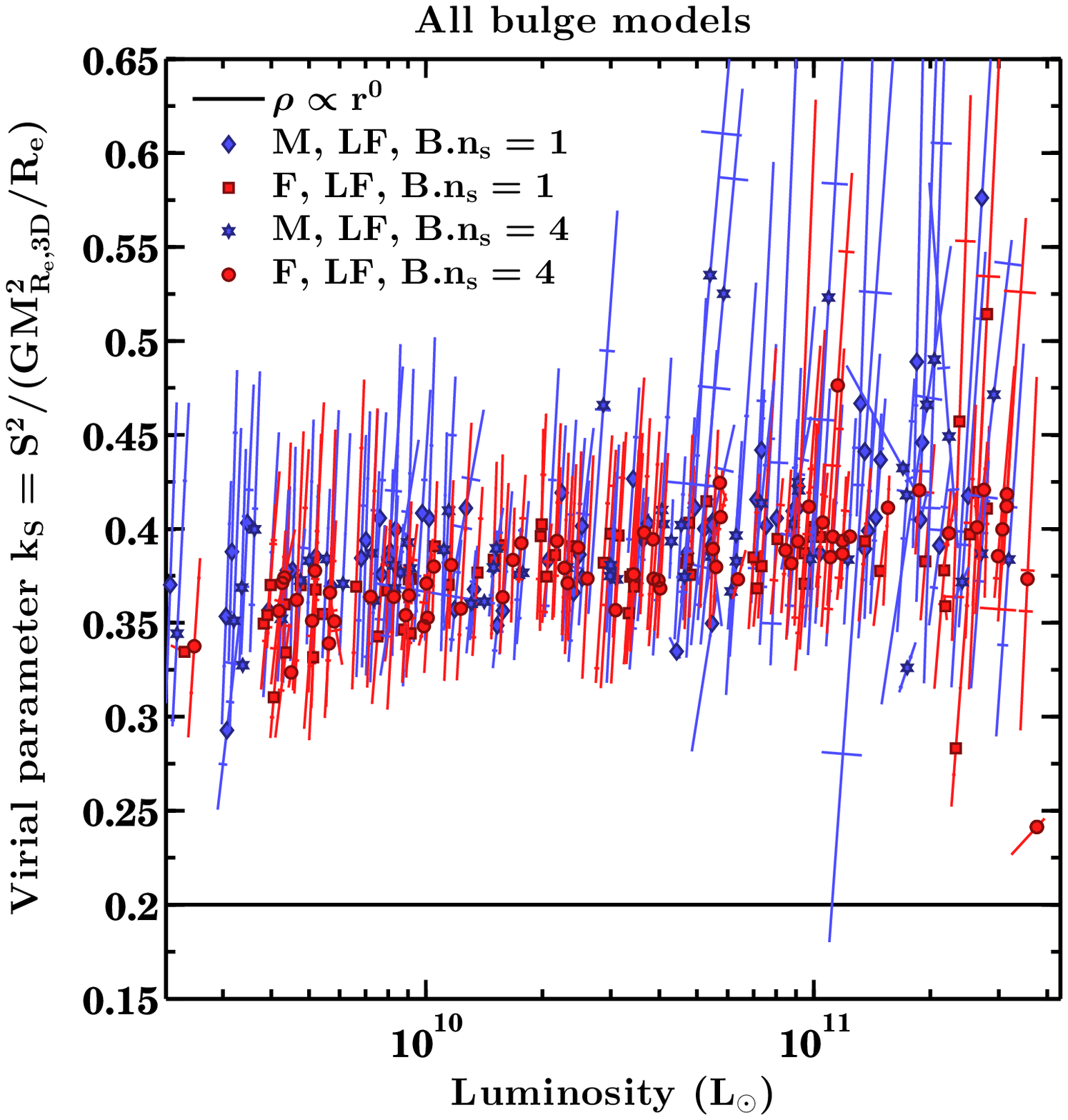}
\includegraphics[width=0.33\textwidth]{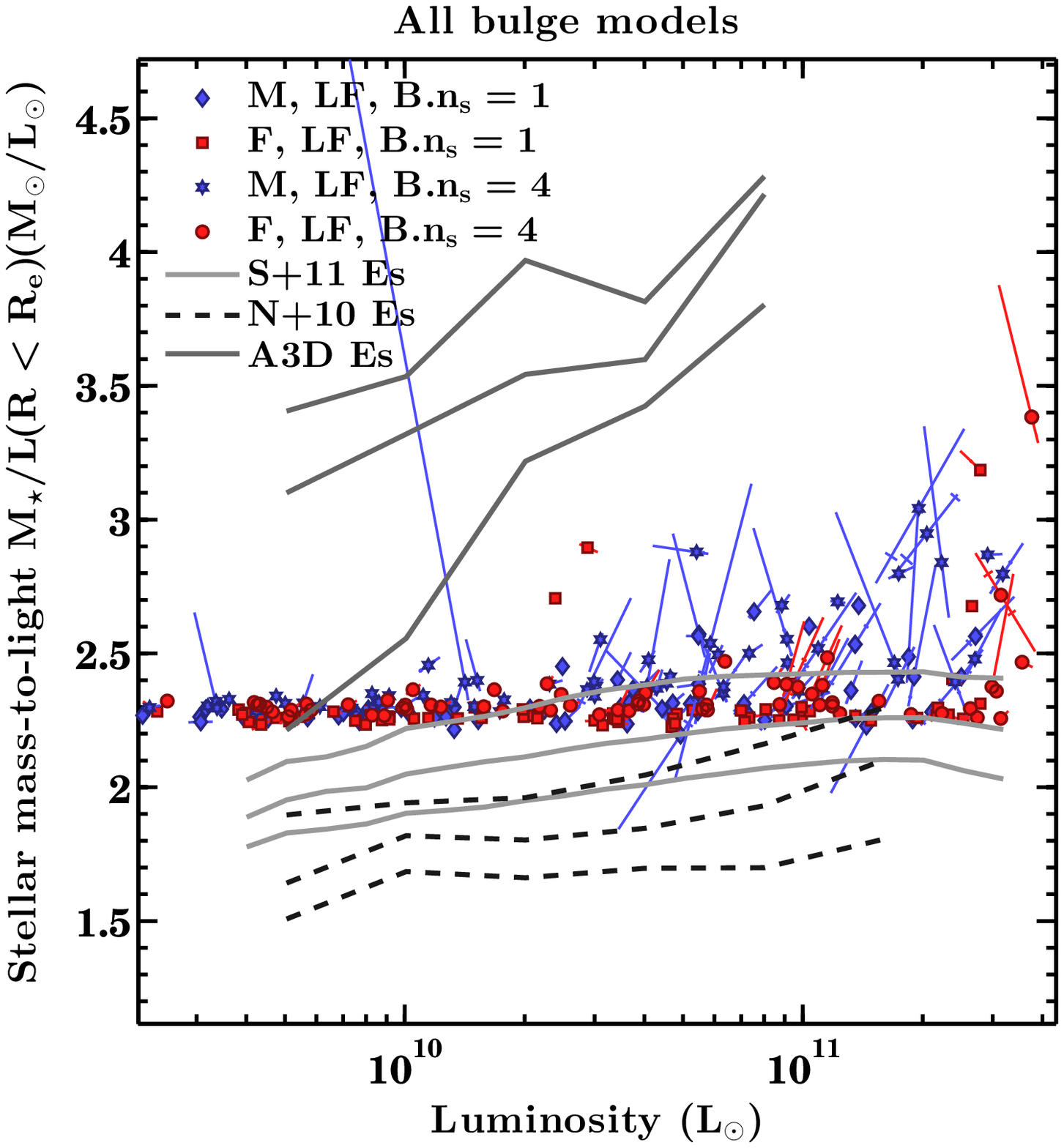}
\includegraphics[width=0.33\textwidth]{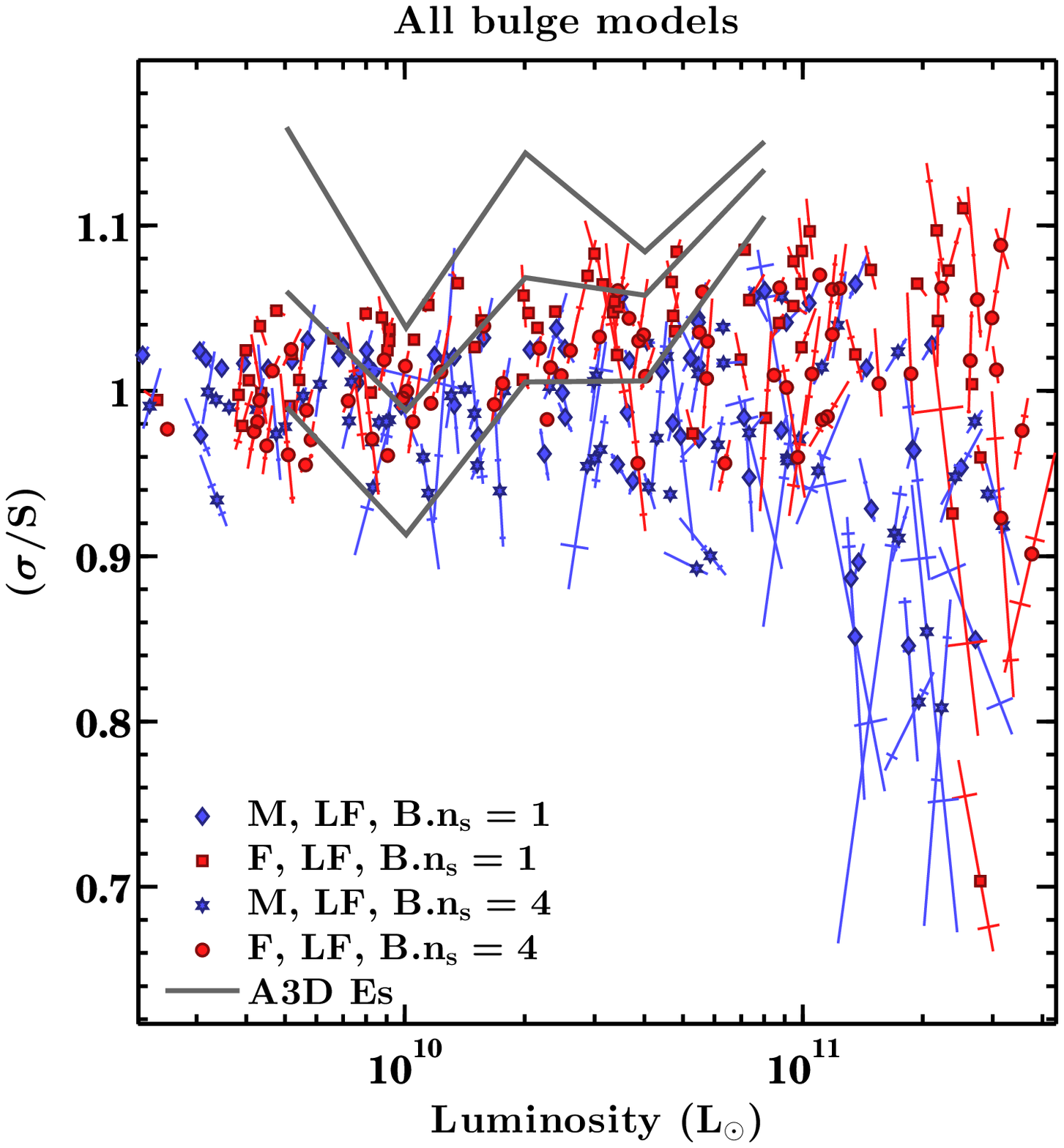}
\caption{Three possible tilt terms $k_{S}$, $M_{\star}/L$ and $\sigma/S$ - the virial parameter, stellar mass-to-light and dispersion to orbital velocity, respectively. Only the virial parameter varies significantly along the simulation FP. Stellar mass-to-light ratios are constant by construction, although some galaxies with substructure have (spuriously) larger ratios. Since most remnants rotate slowly, the $S$ term - $\sigma$ corrected for ordered rotation - is near unity, even for the relatively more rapidly-rotating, faint A3D ellipticals. \label{fig:extratilt}}
\end{figure*}

The three remaining terms that can contribute to the tilt are shown in \figref{extratilt}, beginning with the virial parameter $k_{S}$. For a dispersion-supported, uniform unit sphere, $k_{S} \simeq k = 0.2$, so most of the simulated galaxies have $k_{S}>0.2$ because they are centrally concentrated. For the most part, $k_{S}$ does not vary strongly with any of the FP parameters, although it tends to be larger for more luminous galaxies. At the most luminous end, a number of outliers appear with unusually low values of $k_{S}$. These tend to be systems which were shown to have overestimated $R_{e}$ in Paper I, and their $k_{S}$ values are small only because the $M_{R_{e},3D}^2$ term in the denominator increases more rapidly than $R_{e}$. The outliers with large $k_{S}$ tend to be triaxial systems with strongly projection-dependent $S$.

The remaining two tilt terms are also shown in \figref{extratilt} and contribute little to the tilt in simulations. The $\sigma/S$ term (left panel) is nearly constant at unity in both simulations and in A3D. Again, the main outliers are groups for which $R_{e}$ has been overestimated. Since projected dispersions tend to drop with radius, $\sigma_{e}/\sigma$ becomes significantly smaller than unity if $R_{e}$ is overestimated. While $\sigma_{e}$ is typically quite small in A3D, it is often offset by a substantial rotation correction in $v/\sigma$. This correction is much smaller in the simulated ellipticals, since most are slow rotators (see Paper I).

In the simulations, $M_{\star}/L$ is nearly constant by construction; the only outliers are groups for which a massive satellite appears within $R_{e}$, since we have not masked or fit satellites in the stellar mass maps. By contrast, both observed data sets show a trend of increasing $M_{\star}/L$ with $L$. The A3D ellipticals appear to have a steeper slope, which may be partly due to systematics. A3D $M_{\star}/L$ are derived from fits to spatially-resolved spectra, whereas the SDSS $M_{\star}/L$ is based on fits to broadband colors. The A3D fits also assume a Salpeter IMF rather than the Kroupa IMF used in SDSS; we have adjusted the normalization of the A3D $M_{\star}/L$ by dividing by a factor of 1.7 to correspond to an old, Kroupa IMF, but this normalization may not be self-consistent.

\subsection{The Virial Theorem and the FP}
\label{subsec:virfp}

Having established that the dark matter fraction is likely a major contributor to the tilt, we will now fit various permutations of the FP including each tilt parameter to determine which are closest to the virial FP. To do so, one can fit an FP with $R_{e}$, a velocity ($\sigma$, $\sigma_{e}$ or $S$), and a third parameter other than the usual $L$ or $\mu_{e}$. The values of the parameters are then compared to the expected virial coefficients $a=2$, $b=0.4$ or $\alpha=-2$ and $\beta=1$, depending on whether the third parameter is a mass/luminosity or surface density. In principle, the fits should be convertible using \eqnref{fpcoeff}, but in practice the choice can be significant. As an example, one fit close to the virial plane in the simulations is:
\begin{equation}\label{eqn:virpfit}
\log{R_{e}} = \alpha\log{S} + \beta\log{M_{R_{e},3D}} + \gamma,
\end{equation} 
with $\alpha = -1.97 \pm 0.03$, $\beta = 1.02 \pm 0.01$ and just 0.023 dex scatter. This fit is tabulated as fit \#3 in \tabref{mdyn}, along with the dynamical mass coefficient $c$, which will be discussed further in \subsecref{mdyn}. The fit is almost exactly the virial FP in \eqnref{fpvi}, and should yield parameters of $a = 1.89$, $b = 0.39$. Fitting the same FP with $-2.5\log{(M_{R_{e},3D}/R_{e}^2)}$ as the third parameter yields $a = 1.90 \pm 0.01$, $b = 0.379 \pm 0.003$. These are nearly precisely the expected values, but the errors on $a$ and $b$ are smaller, so that the FP tilt is small but significant. Using $\sigma_{e}$ in place of $S$ gives very similar results, so both are acceptable choices for relatively slowly-rotating galaxies. However, it is not always the case that $a$ and $b$ transform to $\alpha$ and $\beta$ exactly. For example:
\begin{equation}\label{eqn:virpfit2}
\log{R_{e}} = a\log{\sigma} -2.5b\log{(M_{R_{e},3D)}/R_{e}^2} + c,
\end{equation} 

yields $a = 1.94 \pm 0.01$, $b = 0.43 \pm 0.01$, and 0.028 dex scatter. However, fitting a similar FP as in \eqnref{virpfit} returns $\alpha = -1.55 \pm 0.02$, $\beta = 0.90 \pm 0.01$, with 0.025 dex scatter (fit \#1 in \tabref{mdyn}). This is quite far from the virial plane and even more discrepant from the simple conversion using \eqnref{fpcoeff}. It is not generally the case that fits using mass versus mass surface density are identical, and it appears that using $\sigma_{e}$ or $S$ is necessary to obtain a virial FP in both cases.

Even using the \emph{projected} mass $M_{R_{e},2D}$ and $\sigma_{e}$ in \eqnref{virpfit2} returns $a = 1.98 \pm 0.01$ and $b = 0.43 \pm 0.01$, with 0.020 dex scatter; however, the values ($\alpha$,$\beta$) = (-1.65,0.92) are even further from the virial FP (fit \#13 in \tabref{mdyn}). Notably, they also differ from the values quoted by \cite{AugTreBol10} using $\sigma_{e/2}$ and $M_{R_{e}/2}$. These are effectively ($a$,$b$) = (1.83,0.52), or ($\alpha$,$\beta$) = (-1.13,0.81), if transformed with \eqnref{fpcoeff}. It is not clear if systematics or sample selection contribute to the differences, as we have not attempted to match methods with this particular sample or model gravitational lensing.

The one common conclusion from these fits is that most FPs using a total mass of some sort are close to the virial plane; however, reproducing ($\alpha,\beta$) = (-2,1) is more difficult than ($a,b$) = (2,0.4). If the goal is to get as close to the virial FP as possible using both mass and mass surface density, then $R_{e}$, $M_{R_{e},3D}$ and $S$ or $\sigma_{e}$ are appropriate choices for FP variables. 

However, this conclusion is specific to the conditions of our simulated galaxies, and may not apply to galaxies with different formation histories. There is no guarantee that any given mass measure will follow a virial FP. Similarly, even a mass \emph{estimator} that follows a virial FP cannot give an accurate mass without knowing the dynamical mass coefficient (see \subsecref{mdyn}). This is because the virial FP is not, in fact, a trivial consequence of the virial theorem. Although every merger remnant becomes very nearly virialized with a virial ratio of $1 \pm 1$\%, biased subsets of a virialized galaxy need not have unity virial ratios. The local virial ratio of the total mass within $R_{e},3D$ in the merger remnants ranges from 0.4--0.5. This is expected, because the interior mass of a galaxy (stellar or total) is a very biased subset, residing in the deepest part of the potential well. Likewise, stars are biased tracers of the potential of any galaxy with an extended dark halo. In principle, any variation of the local virial ratio within a given size measure $R$ can translate into variation of the virial parameter $k$ and hence a tilt in the FP. Only a homologous population of galaxies can be expected to have constant local virial ratios within $R$, while the value of the local virial ratio (which is reflected in the dynamical mass coefficient) should be less than unity.

\subsection{Dynamical Masses}
\label{subsec:mdyn}

The existence of a tight mass FP with the virial coefficients implies that one can extract a dynamical mass of the form $M_{dyn}=c\sigma^{2}R/G$ as long as $c$ is known. In \subsecref{virfp}, we showed that \eqnref{virpfit} is very nearly a virial FP, and so one should be able to derive an accurate value for $M_{R_{e},3D}$ from $S$ and $R_{e}$ alone. However, the precise value of $c$ depends on the structure of the galaxy, while the constancy of $c$ relies on the assumption of homology. We already showed in \figref{extratilt} (left panel) that the virial parameter is not exactly constant and has significant projected-dependent scatter. We also demonstrated in the previous section that stellar virial ratios and virial ratios within $R_{e}$ are neither exactly constant nor unity, so these assumptions must be tested.

\begin{figure*}
\includegraphics[width=0.33\textwidth]{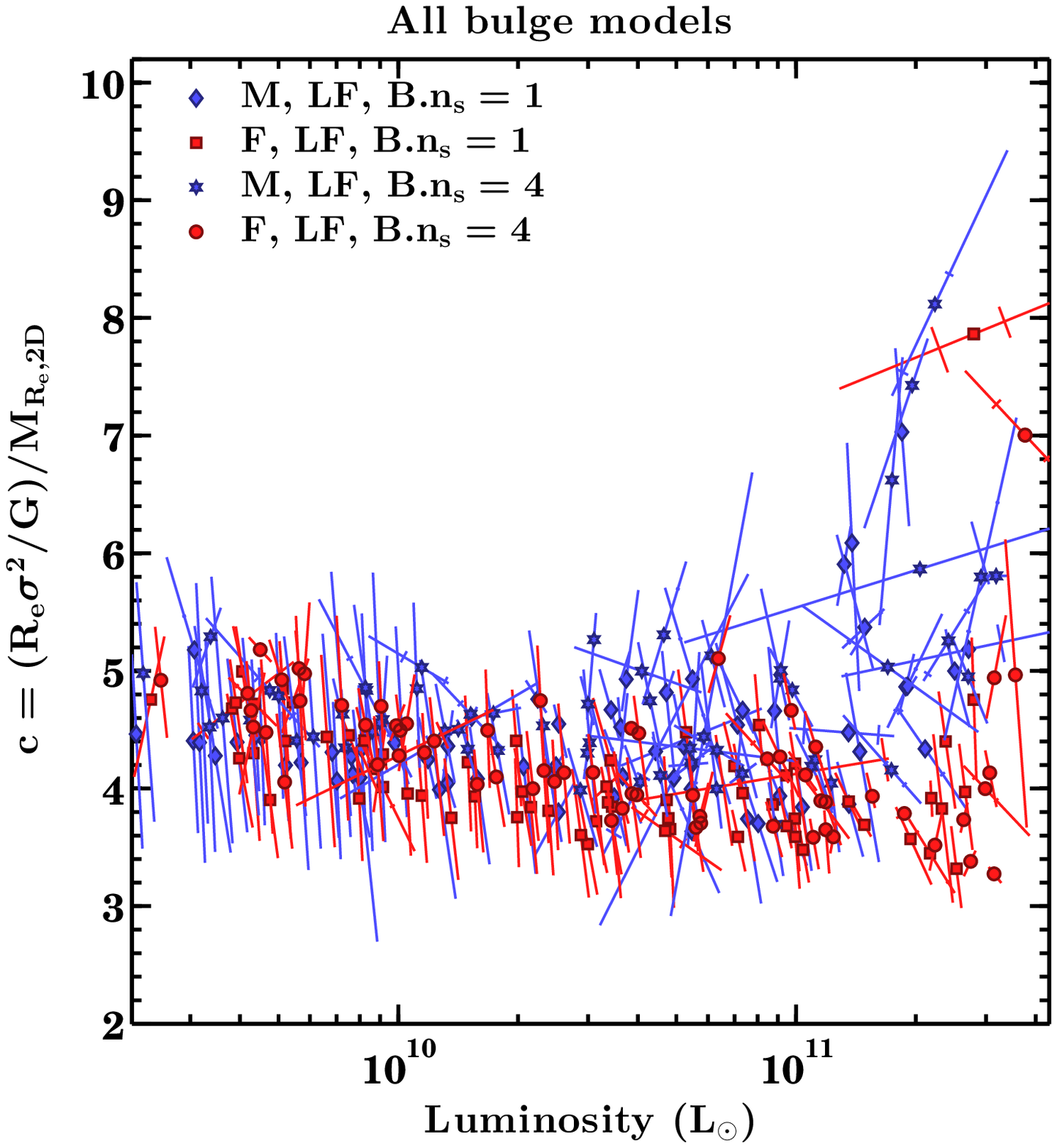}
\includegraphics[width=0.33\textwidth]{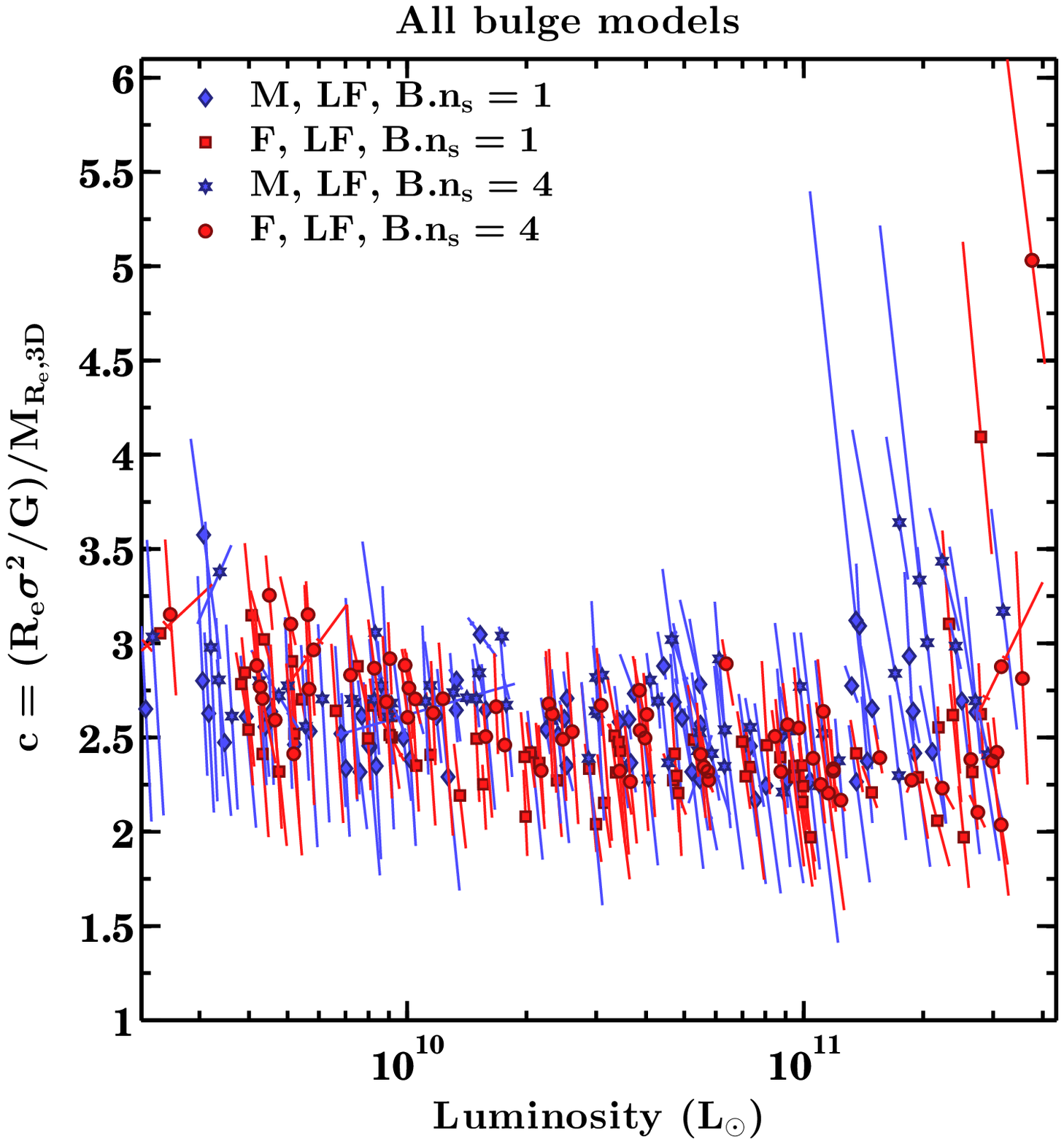}
\includegraphics[width=0.33\textwidth]{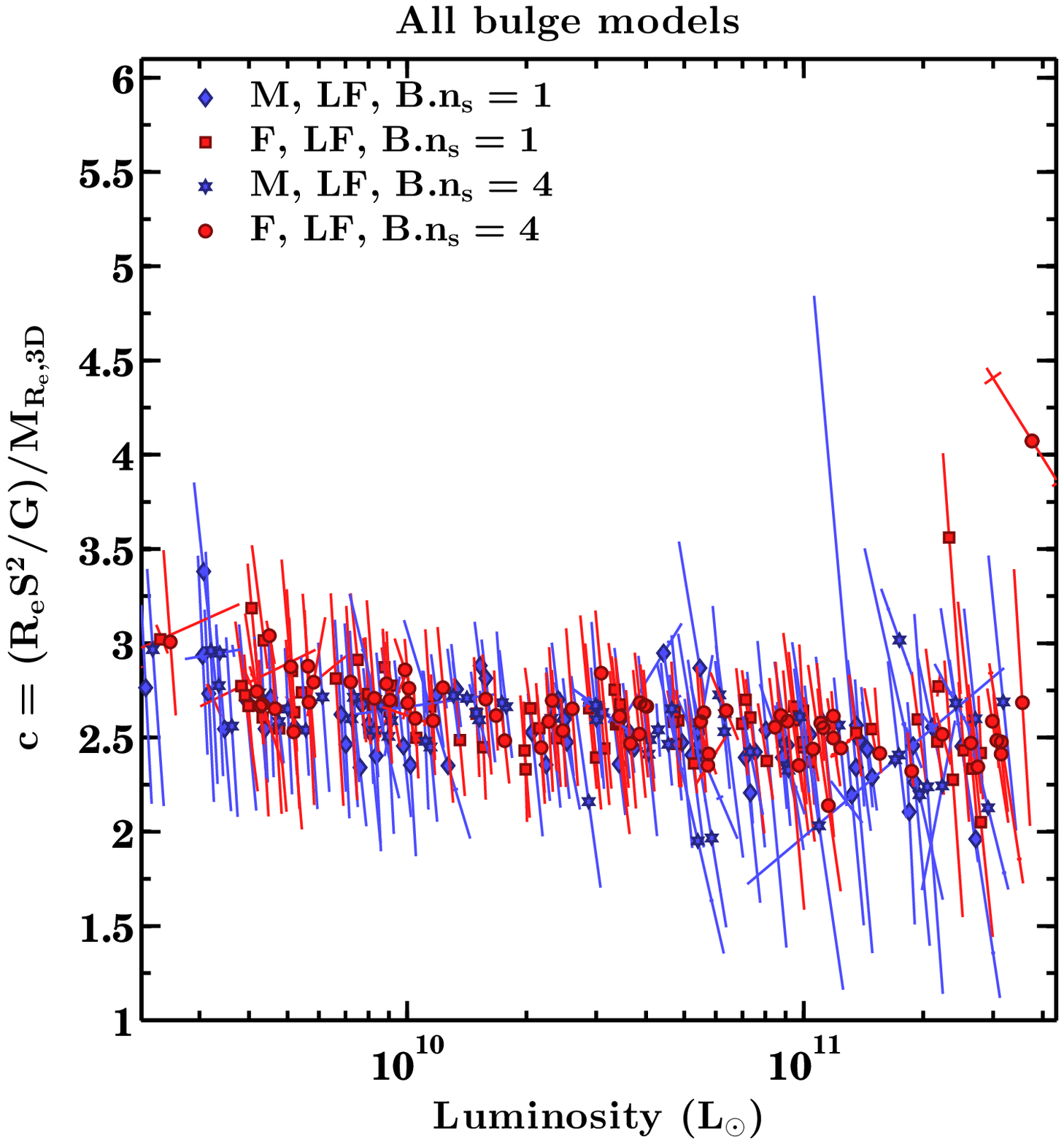}
\caption{Values of $c$, the dynamical mass coefficient (or the dimensional-to-total mass ratio), for various definitions of the dynamical and total mass. In the two leftmost panels, $M_{dyn}=c\sigma^{2}R_{e}/G$; in the right panel, $S$ is used in place of $\sigma$. The left panel ratio is to the projected mass, $M_{R_{e},2D}$, while the two rightmost panels use $M_{R_{e},3D}$.
\label{fig:mdyn}}
\end{figure*}

Values of $c$ for a variety of dynamical mass estimators are shown in \figref{mdyn}. For the projected mass $M_{R_{e},2D}$, values of $c$=3.5--5 are appropriate when using $\sigma$ and $R_{e}$; for $M_{R_{e},3D}$, $c$=2.5--3 is more suitable. The tightest correlations are found by replacing $\sigma$. Using $S$ gives a median $c$ of 2.58 and 0.06 dex scatter, while $\sigma_{e}$ produces $c$ = 2.63 with 0.05 dex scatter. As \figref{mdyn} demonstrates, using $S$ lowers the projection-dependent scatter but cannot remove it entirely - more extreme projections require values of $c$ as small as 1.5 or as large as 3.5. While the correlation with $R_{e}$ and $S$ is fairly tight, there are several notable outliers (again, in galaxies where $R_{e}$ is overestimated) and a hint of a shallow, luminosity-dependent slope in $c$. Nonetheless, for a typical random projection, $R_{e}$ and $S$ or $\sigma_{e}$ can be used to estimate $M_{R_{e},3D}$ to within 10--15\%. at least partly due to the fact that these three variables define a tight virial FP with 0.02 dex scatter.

Other dynamical mass definitions have been proposed in the literature. On theoretical grounds, \cite{WolMarBul10} advocate the use of $M_{1/2}$, where $M_{1/2}$ is the total mass within $r_{1/2}$, and $r_{1/2}$ is the 3D radius containing half of the total luminosity of the galaxy. Using Jeans models, they derive a relation $M_{1/2} = 3r_{1/2}\sigma^2$, arguing that $r_{1/2}$ is a unique radius at which the effects of anisotropy are minimized. For a Sersic profile with 0.5 $<n_{s}<$ 10, $r_{1/2}=1.34R_{e}$ is an excellent approximation \citep{Cio91}, and so one can also write $M_{1/2} = 4R_{e}\sigma^2$ without having to infer $r_{1/2}$.

Another alternative dynamical mass was defined by \citet{CapScoAla13}. They applied a variant of Jeans modelling (``JAM'' models) to A3D galaxies to obtain a mass $M_{JAM}$, where $M_{JAM} = L[(M/L)_{JAM}(r < R_{e})]$. Their Figure 14 suggests that $M_{JAM}/M_{vir}$ shows minimal scatter (0.08 dex), provided that $M_{vir}=3.9R_{e,maj}\sigma_{e}^2$, where $R_{e,maj}$ is the major axis (not circularized) $R_{e}$. Similarly, \citet{CapScoAla13} argue that $M_{JAM}$ is an appropriate dynamical mass, because it defines a tight virial FP with $R_{e,maj}$ and $\sigma_{e}$ (see their Figure 12).

We have tested both of these dynamical mass estimators, listing the values of $c$ and the mass FP parameters $\alpha$ and $\beta$ for each estimator in \tabref{mdyn}. $r_{1/2}$ is measured as the radius containing half of the stellar mass within $r_{200}$. This is equivalent to the half-luminosity radius, since $M_{\star}/L$ is nearly constant, and for most galaxies, almost all of the bound star particles lie within $r_{200}$. The median $c$ for $M_{1/2}$, $r_{1/2}$ and $\sigma$ (fit \#4 in \tabref{mdyn}) is 2.67, with 0.08 dex scatter. This is within 10\% of the value of 3 quoted by \citet{WolMarBul10}, but the scatter with $M_{1/2}$ is larger than for $M_{R_{e},3D}$. The scatter is lower if using $\sigma_{e}$, at 0.06 (fit \#5), while the median of 2.74 remains lower than 3. 

\begin{table*}
\centering
\caption{Mass Fundamental Plane and Dynamical Mass Estimators}
Simulations: All B.$\mathrm{n_{s}}$, ten equally-spaced, randomly-oriented projections \\
\begin{tabular}{cccccccccc}
\hline 
\# & $M$ & $R$ & $K$ & $c$ & $\Delta c$ & $\alpha$ & $\beta$ & $\gamma$ & R.M.S. \\
\hline
1 & $M_{R_{e},3D}$ & $R_{e}$ & $\sigma$ & 2.550 $\pm$ 0.010 & 0.065 & -1.553 $\pm$ 0.022 & 0.899 $\pm$ 0.007 & -5.640 $\pm$ 0.030 & 0.024 \\
2 & $M_{R_{e},3D}$ & $R_{e}$ & $\sigma_{e}$ & 2.626 $\pm$ 0.006 & 0.053 & -1.979 $\pm$ 0.032 & 1.012 $\pm$ 0.009 & -5.957 $\pm$ 0.036 & 0.021 \\ 
3 & $M_{R_{e},3D}$ & $R_{e}$ & $S$ & 2.575 $\pm$ 0.006 & 0.058 & -1.969 $\pm$ 0.031 & 1.017 $\pm$ 0.009 & -6.016 $\pm$ 0.038 & 0.023 \\
4 & $M_{1/2}$ & $r_{1/2}$ & $\sigma$ & 2.669 $\pm$ 0.013 & 0.080 & -1.220 $\pm$ 0.019 & 0.803 $\pm$ 0.005 & -5.269 $\pm$ 0.024 & 0.027 \\ 
5 & $M_{1/2}$ & $r_{1/2}$ & $\sigma_{e}$ & 2.741 $\pm$ 0.008 & 0.061 & -1.548 $\pm$ 0.022 & 0.886 $\pm$ 0.006 & -5.498 $\pm$ 0.024 & 0.023 \\
6 & $M_{1/2}$ & $r_{1/2}$ & $S$ & 2.676 $\pm$ 0.012 & 0.065 & -1.535 $\pm$ 0.023 & 0.888 $\pm$ 0.006 & -5.539 $\pm$ 0.026 & 0.025 \\
7 & $M_{1/2}$ & $R_{e}$ & $\sigma$ & 5.045 $\pm$ 0.027 & 0.141 & -3.764 $\pm$ 0.189 & 1.390 $\pm$ 0.045 & -6.665 $\pm$ 0.119 & 0.052 \\
8 & $M_{model}$ & $R_{e,maj}$ & $\sigma$ & 5.295 $\pm$ 0.016 & 0.061 & -1.918 $\pm$ 0.022 & 1.007 $\pm$ 0.007 & -6.342 $\pm$ 0.032 & 0.024 \\ 
9 & $M_{model}$ & $R_{e,maj}$ & $\sigma_{e}$ & 5.447 $\pm$ 0.013 & 0.060 & -2.509 $\pm$ 0.037 & 1.167 $\pm$ 0.011 & -6.889 $\pm$ 0.046 & 0.021 \\
10 & $M_{model}$ & $R_{e,maj}$ & $S$ & 5.367 $\pm$ 0.014 & 0.067 & -2.509 $\pm$ 0.038 & 1.176 $\pm$ 0.011 & -6.974 $\pm$ 0.051 & 0.024 \\
11 & $M_{model}$ & $R_{e}$ & $\sigma$ & 6.296 $\pm$ 0.018 & 0.059 & -1.873 $\pm$ 0.020 & 1.013 $\pm$ 0.006 & -6.573 $\pm$ 0.029 & 0.021 \\
12 & $M_{R_{e},2D}$ & $R_{e}$ & $\sigma$ & 4.272 $\pm$ 0.019 & 0.068 & -1.319 $\pm$ 0.013 & 0.829 $\pm$ 0.004 & -5.568 $\pm$ 0.020 & 0.021 \\
13 & $M_{R_{e},2D}$ & $R_{e}$ & $\sigma_{e}$ & 4.438 $\pm$ 0.011 & 0.045 & -1.650 $\pm$ 0.017 & 0.916 $\pm$ 0.005 & -5.826 $\pm$ 0.020 & 0.017 \\
14 & $M_{200}$ & $r_{1/2}$ & $\sigma$ & 53.46 $\pm$ 0.32 & 0.107 & -2.938 $\pm$ 0.053 & 1.382 $\pm$ 0.016 & -9.819 $\pm$ 0.087 & 0.029 \\
15 & $M_{group}$ & $r_{1/2}$ & $\sigma$ & 68.37 $\pm$ 0.34 & 0.096 & -2.783 $\pm$ 0.047 & 1.315 $\pm$ 0.014 & -9.462 $\pm$ 0.078 & 0.028 \\
\hline
\end{tabular}
\tablecomments{Each dynamical mass estimator is based on a mass ($M$), radius ($R$), and kinematic ($K$) tracer. For each estimator, the median $c$ is tabulated, from $c=(RK^{2}/G)/M$, along with the error on the median and the scatter $\Delta c$ about the median. Also shown is the best-fit $\log{R}=\alpha\log{K} + \beta\log{M} + \gamma$, with the r.m.s. scatter about this relation.
\label{tab:mdyn}}
\end{table*}

Using $R_{e}$ in place of $r_{1/2}$ gives a median $c$ = 5.05 with 0.14 dex scatter (fit \#7 in \tabref{mdyn}). This value of $c$ is a full 25\% larger than the nominal value of 4. Furthermore, the best-fit FP is wildly different from the virial FP, and has much larger scatter than any other mass FP in \tabref{mdyn} (using $\sigma_{e}$ or $S$ does not improve either fit). This is mainly because the approximation $r_{1/2}=1.34R_{e}$ does not hold for our galaxies - instead, $r_{1/2}\approx 1.88R_{e}$, with 0.13 dex scatter. This limits the usefulness of $M_{1/2}$, since $r_{1/2}$ is not directly measurable. The tight relation between $R_{e}$ and $r_{1/2}$ only applies for pure, spherical Sersic profiles. Realistic merger remnants and ellipticals alike are not perfectly spherical. Systematic effects in SDSS-quality imaging limit the ability to recover $R_{e}$ and $n_{s}$, even if ellipticals have perfect Sersic profiles, while our merger remnants evidently only approximately follow Sersic profiles.

We define an analogous mass to $M_{JAM}$ as $M_{model}=L[M/L(r < R_{e,3D})]$, but using the actual $M/L$ within $R_{e,3D}$. $M_{JAM}$ equals $M_{model}$ if the JAM model derives the correct $M/L$. Note that $M_{model}$ does not really have a physical meaning, as it is a projected luminosity multiplied by a mass-to-light ratio within a sphere. If $R_{e}$ is correct and truly the half-light radius, \emph{and} if $M/L(r<R_{e,3D})=(M/L)(r<r_{1/2})$, then $M_{model} = 2 M_{1/2}$. In practice, $M/L$ should increase with radius, and so $M_{model}/2$ is really a lower limit on $M_{1/2}$.

Fit \#9 in \tabref{mdyn}, using $M_{model}$, $R_{e,maj}$ and $\sigma_{e}$, results in a median $c$ = 5.45 with 0.06 dex scatter (note that $c$ = 2.72 for $M_{model}$/2). $M_{model}$ can be recovered nearly as accurately as $M_{R_{e},3D}$; however, the value of $c$ is considerably larger than the value of 3.9 quoted by \citet{CapScoAla13}. This is mainly due to systematics - the values of $R_{e}$ used by \cite{CapScoAla13} are not derived from Sersic profile fits and tend to be larger than those from 2D, single Sersic GALFIT fits. Perhaps for similar reasons, we are also unable to recover a tight virial FP using $M_{model}$. Our best-fit relation for $M_{model}$ is $R_{e,maj} \propto \sigma_{e}^{-2.51 \pm 0.04}M_{model}^{1.17 \pm 0.01}$ (fit \#9 again).

We conclude that $M_{R_{e},3D}$ is the only dynamical mass that satisfies two broad conditions. First, it can be accurately recovered within 10--15\% using $R_{e}$ and $S$ or $\sigma_{e}$. Secondly, it is linked to the fundamental plane, in that $M_{R_{e},3D}$, $S$ and $R_{e}$ define a nearly exact virial plane (\eqnref{virpfit}). This latter criterion was advocated by \citet{CapScoAla13}, but using $M_{model}$; we do not confirm this result using $M_{model}$ and $R_{e,maj}$ based on 2D Sersic fits. $M_{model}$ still defines a tight mass FP, just not with exactly the virial scalings - unless $\sigma$ is used in place of $\sigma_{e}$, in which case fit \#8 comes close.

Still, a nagging question remains - does the fact that $M_{R_{e},3D}$, $R_{e}$ and $S$ (or $\sigma_{e}$) define a virial FP hold any fundamental significance? Similarly, can the value $c\approx$2.6 be derived from the SVT? To answer this, we begin with the virial ratio $q_{50} = -2T_{R_{e},3D}/W_{R_{e},3D}$, and note that $q_{50} \approx 0.45$ in typical merger remnants. We emphasize that $q_{50}$ depends on the dynamics of the stars and halo, generally being lower than unity, and must not be assumed to equal unity. Since $q_{50}$ is not constant, it is not surprising that $M_{dyn}/M_{R_{e,3D}}$ varies slightly too.

Next, we need relations between kinetic and potential energies within $R_{e}$ and observables. Let $q_{W,50}=W_{R_{e},3D}/(GM^2/R_{e})$; $q_{W,50} \approx 3.0$ in the simulations. Then if $S$ traces the total kinetic energy, $2T_{50} = 3\sigma^2 M (S/\sigma)^2$, where $S/\sigma \approx 0.95$ (\figref{extratilt}). Inserting these values into the virial ratio yields:
\begin{equation}
M = [3(S/\sigma)^2/(q_{50}q_{W,50})]\sigma^2R_{e}/G.
\end{equation}
Thus, the dynamical mass coefficient should be $c=3(S/\sigma)^2/(q_{50}q_{W,50})$. For typical simulations, this is $3(0.95)^2/(0.45 \times 3.0)=2.0$. This is somewhat lower than the value of 2.6 shown in \figref{mdyn}. The main reason for this is that $\langle \sigma_{DM}/\sigma \rangle=1.34$ in the simulations, and so both $\sigma$ and $S$ underestimate $T$, which includes the dark matter kinetic energy. This is irrelevant if one simply uses an empirical value of $c$, but important if one wishes to derive a true virial mass using the SVT. If one is not really deriving a virial mass using the SVT, then it is not clear why one should favor a mass estimator that is best fit by the virial FP in the first place. In principle, any correlation will do, even if it is not strictly a dimensional mass defined by a virial FP.

Finally, having gone through this exercise, it is worth pointing out that none of these dynamical masses are even remotely close to the total mass of the galaxy. Using $r_{1/2}$ and $\sigma$ to estimate $M_{200}$ (including substructure), we obtain $c$=53.5 with 0.11 dex scatter (fit \#14 in \tabref{mdyn}); using $R_{e}$ yields still larger $c$ and scatter. The dynamical mass estimators we have discussed here are mainly useful for setting limits on the dark matter content within elliptical galaxies, given rather strict assumptions.

\subsection{Consistency Check}
\label{subsec:modelcheck}

As we have shown, most of the tilt of the FP in the simulated galaxies is caused by varying dark matter fractions, and while the virial parameter $k$ (or $k_{S}$) is not exactly constant, it does not contribute much to the tilt. We check the consistency of this result by creating a mock fundamental plane from simple, spherical bulge plus halo systems, using the same GalactICS code \citep{WidDub05,WidPymDub08} as was used to generate the model spirals in Paper I. 

Each galaxy consists of a Sersic profile bulge and a dark halo. The stellar component follows a scaling relation of $R_{e} \propto M_{\star}^{0.7}$, with $\sigma$ scaled to create a virial FP and $R_{e}$ and $\sigma$ normalized to followed the observed relations for ellipticals in Paper I. This results in a scaling $\sigma^2 \propto M/R \propto M^{0.3}$. The halo scale radius $r_{s}=2R_{e}$, and the scale velocity $v=1.2\sigma$, roughly consistent with the simulated ellipticals in this paper. The dark halo has an inner density profile of $\rho \propto r^{-1}$, an outer profile of $\rho \propto r^{-2.5}$, and truncates smoothly to $\rho=0$ from $30r_{s}$ to $35r_{s}$. This is much like a truncated NFW profile \citep{NavFreWhi97}. The total halo mass is $37M_{\star}$, equivalent to assuming that $\Omega_{\star}<0.01$.

We create 49 such galaxies with bulge $n_{s}=4$, $M_{\star}/L=2$, and $R_{e}$ ranging from 1.4 to 27 kpc. Modest intrinsic scatter is added by making some galaxies slightly over- or under-massive for their size. We then generate mock images using the same methodology as Paper I; the FP for the galaxies has $a=2.06 \pm 0.02$, $b=0.42$ and negligible scatter. The small deviation from the virial plane is entirely due to systematics. Direct, one-dimensional fits to the surface brightness profile without a variable sky background give $(a,b) = (2.00 \pm 0.01,0.39)$.

We then repeat this process, introducing non-homology by scaling $n_{s}$ by a slope of 1 per dex in $M_{\star}$, roughly consistent with the observed and simulated relations in Paper I. Each galaxy keeps the same $R_{e}$, $M_{\star}$ and $M$. Since $n_{s}>4$ profiles are more centrally concentrated, $\sigma$ is larger than in the $n_{s}=4$ case (or smaller if $n_{s}<4$), and so $\sigma$ scales more steeply with $M_{\star}$. For this sample, we find $(a,b) = (1.94 \pm 0.04, 0.41 \pm 0.01)$ - as expected, the $a$ parameter is lower, since $\sigma$ scales more steeply with $R_{e}$ as well. However, $a$ is not significantly smaller than 2, and while the difference between $a$ in this variable $n_{s}$ sample to the $n_{s}=4$ sample is marginally significant, it is mainly systematic. One-dimensional fits give $(a,b) = (1.98 \pm 0.01,0.40 \pm 0.01)$, at best a barely significant change from the $n_{s}=4$ case.

In conclusion, the effect of structural non-homology in galaxies with Sersic stellar mass profiles and fixed halo profiles is negligible. However, this may not necessarily be the case if the stellar profile deviates from a Sersic law, or if the dark halo profiles scale differently from the stellar profiles. For example, one could induce non-homology with otherwise self-similar dark halos if $d\log{R_{s}}/d\log{R_{e}} > 0$, which would result in larger dark matter fractions in larger galaxies \citep[see][who conducted a similar exercise]{BorSalDan03}. Unfortunately, this interpretation is overly simplistic - the halos in the simulated galaxies are not completely self-similar, and for the common profiles we have fit, $d\log{R_{s}}/d\log{R_{e}} < 0$ and the central dark matter density $\rho_{0,DM}$ is not constant either. We leave further analysis of dark halo profiles to simulations with fully cosmological merger histories and halo profiles.

\section{An Alternative Formulation of the Tilt}
\label{sec:alttilt}

Only the total kinetic energy $T$ and potential $W$ obey the SVT in virialized galaxies. Given that, one can derive the FP by re-writing the virial parameter $k=\sigma^2R/(GM)$ as the ratio of two dimensionless parameters:
\begin{equation}\label{eqn:qs}
q_{T} = M_{\star}\sigma^{2}/2T \textrm{ and } q_{W} = -(GM_{\star}^2/R)/W,
\end{equation}
such that $k = (q_{T}/q_{W})(-2T/W)$, or simply $k=q_{T}/q_{W}$ if the galaxy is virialized. Then the virial plane becomes:
\begin{equation}\label{eqn:fpva}
\log{R} = -2\log{\sigma} + \log{M_{\star}} + \log{G} + \log q_{T} - \log q_{W}.
\end{equation}
This form is virtually the same as \eqnref{fpv2} if one adds the tilt term $\log(M_{\star}/L)$ and interchanges $k$ and $q_{T}/q_{W}$. The purpose of the latter substitution is to include the virial ratio $-2T/W$ into two separable tilt terms based on physically meaningful (though not easily measurable) parameters. To illustrate, we use $R=R_{e}$ and $M=M_{R_{e},3D}$ to fit the equation: 
\begin{equation}\label{eqn:}
\log{R_{e}} = a\log{L} + b\log{\sigma} + d\log{q_{T}} + e\log{q_{W}} + f\log{\Upsilon_{\star}} + z,
\end{equation}
yielding $a=-2.04 \pm 0.01$, $b=1.01$, $d=1.02$, $e=-1.02$, $f=0.92 \pm 0.02$ and $z=-5.33 \pm 0.01$, with 0.002 dex scatter. This is nearly exactly the virial plane with 0.4\% scatter, except that the constant $z$ is insignificantly larger than $\log G=-5.37$ and the $\Upsilon_{\star}$ term is underfit (likely because the stellar mass maps are not satellite-subtracted). This is in spite of the fact that we have not accounted for any systematics in our definition of $R$ and $M$, and that there is nothing inherently special about the choice of $R_{e}$ as a size.

\begin{figure*}
\includegraphics[width=0.325\textwidth]{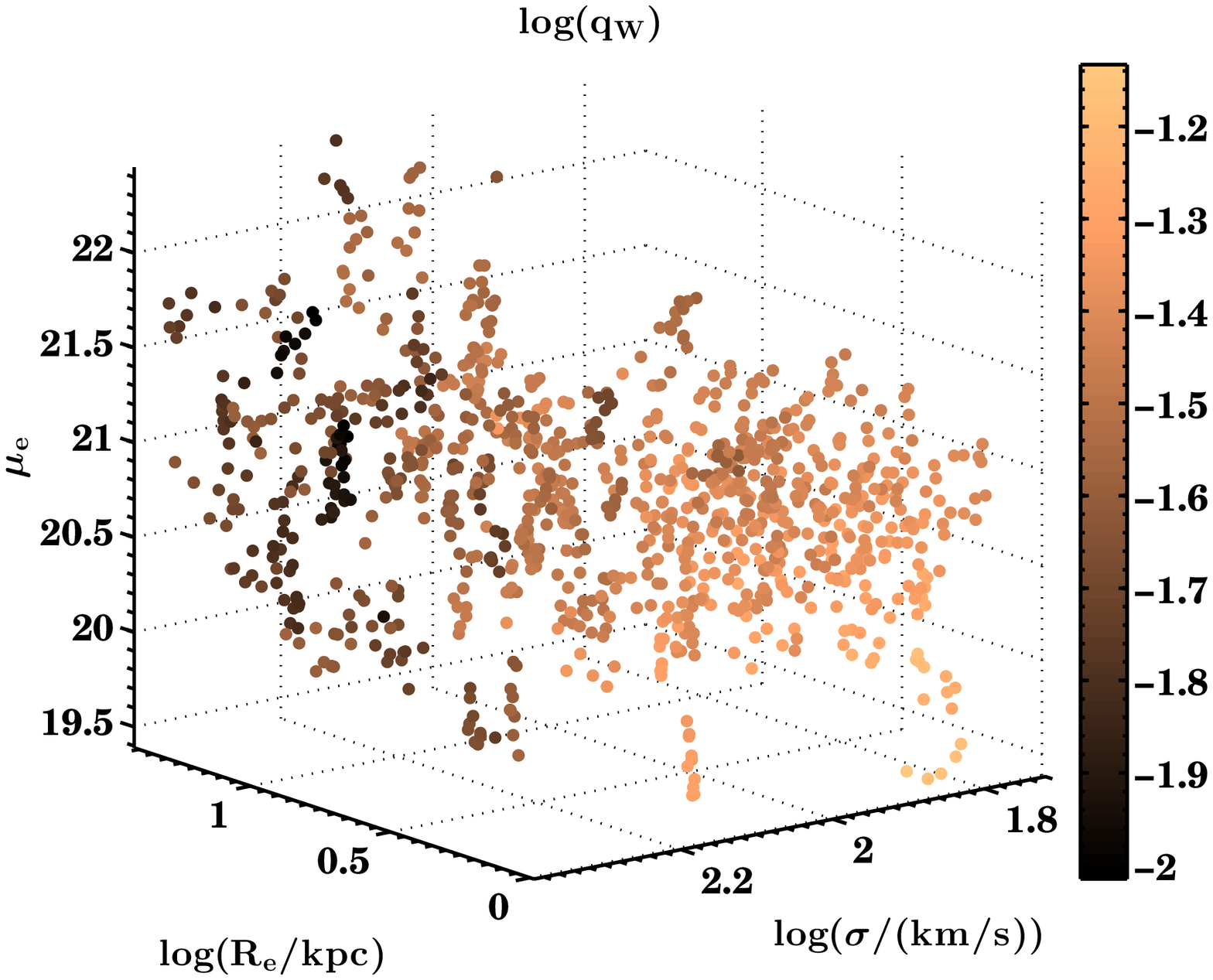}
\includegraphics[width=0.325\textwidth]{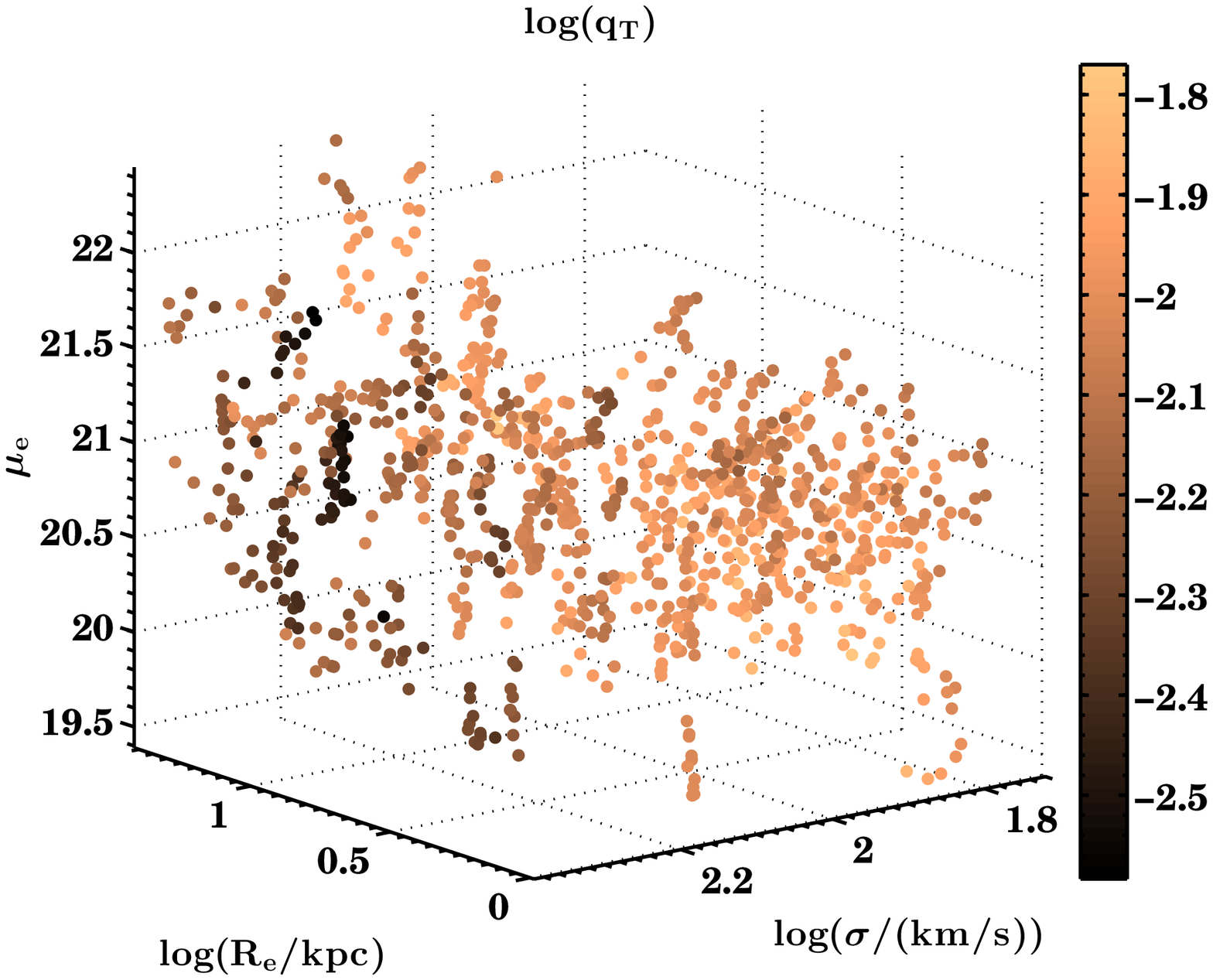}
\includegraphics[width=0.325\textwidth]{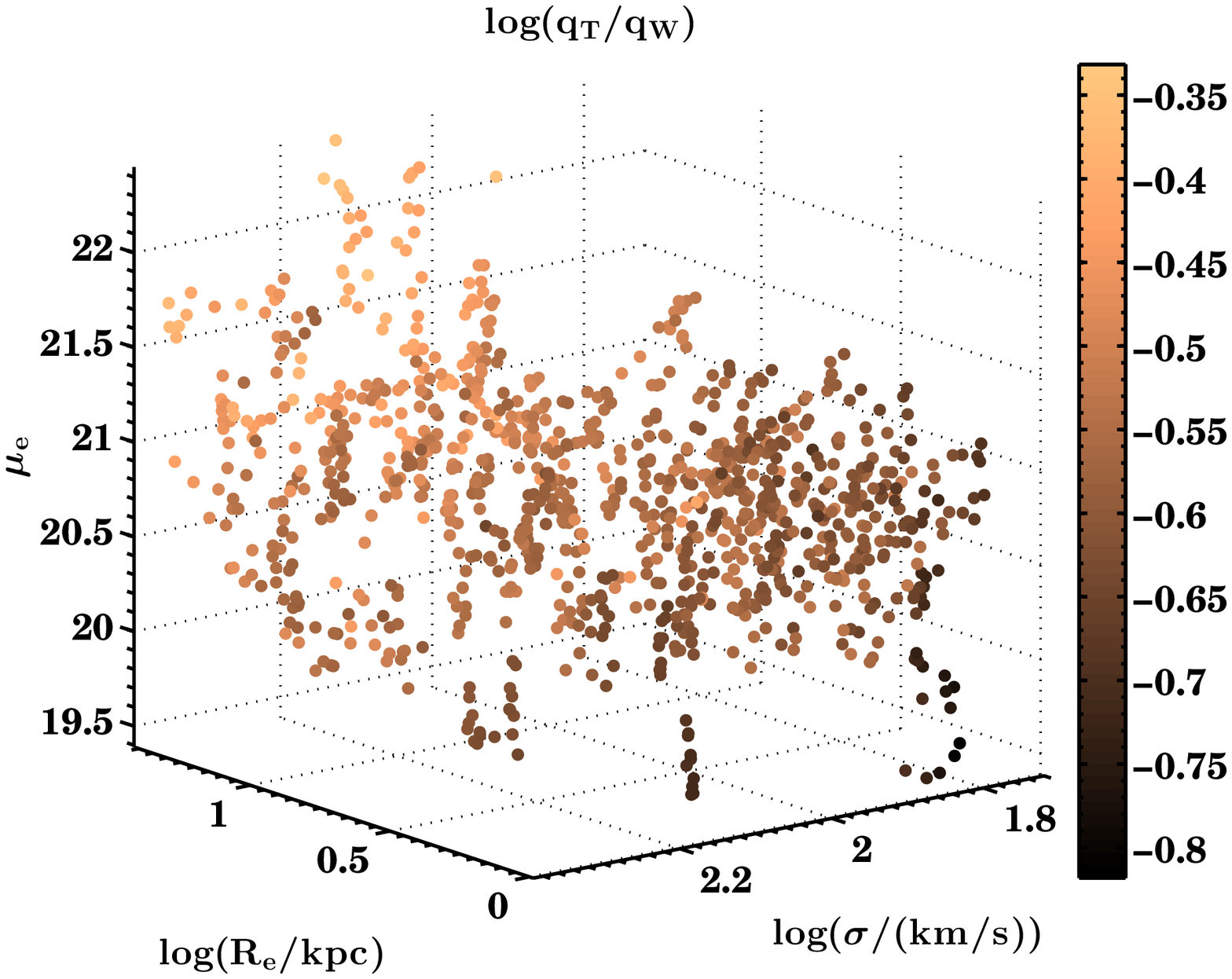}
\caption{The two tilt terms $q_{T}$, $q_{W}$ and the net tilt $q_{T}/q_{W}$ as a function of the three fundamental plane parameters, shown roughly face-on.
\label{fig:q}}
\end{figure*}

The two tilt terms and their ratio $q_{T}/q_{W}$ are shown in a 3D plot in \figref{q}. Both terms clearly vary across the plane, although neither term produces a clean gradient individually. However, the total tilt $q_{T}/q_{W}$ does vary smoothly across the plane, which is not entirely surprising given how strongly $q_{T}$ and $q_{W}$ are correlated: $\log{q_{T}} = 0.86\log{q_{W}} + 1.40$ with 0.04 dex scatter. 

To make the use of these terms more explicit, we split each into two components: 
\begin{equation}\label{eqn:qt}
q_{T} = (T_{\star}/T)M_{\star}\sigma^{2}/2T_{\star},
\end{equation}

The $M_{\star}\sigma^{2}/2T_{\star}$ term is largely a nuisance parameter - it is equal to 1/3 in a spherical, isotropic, dispersion-supported system, but can vary for more complex systems. Now $M_{\star}/M$ is the familiar stellar mass fraction term, joined by a very similar ratio in $T_{\star}/T$, which encompasses non-homology in the stellar kinetic energy fraction. The second tilt term $q_{W}$ can also be decomposed: 
\begin{equation}\label{eqn:qw}
q_{W} = [(-GM^{2}/R)/W](M_{\star}/M)^{2}.
\end{equation}

The dark matter fraction returns as a tilt term, with another non-homology term: $(-GM^{2}/R)/W$, which includes structural non-homology in the total mass profile, rather than kinematics. This latter term is, like $T_{\star}/T$, virtually impossible to measure observationally, but we are free to measure both in the simulations.

\section{The Origin of the Tilt}
\label{sec:origin}

We have just shown that the tilt of the stellar mass FP can be restated as originating from three key terms - $M_{\star}/M$,  $T_{\star}/T$ and $(-GM^{2}/R)/W$. The first term, the stellar mass fraction, has already been shown to be a major contributor, while the last term is approximately constant at a value of 3 in most of the remnants. What about $T_{\star}/T$?

\begin{figure*}
\includegraphics[width=0.50\textwidth]{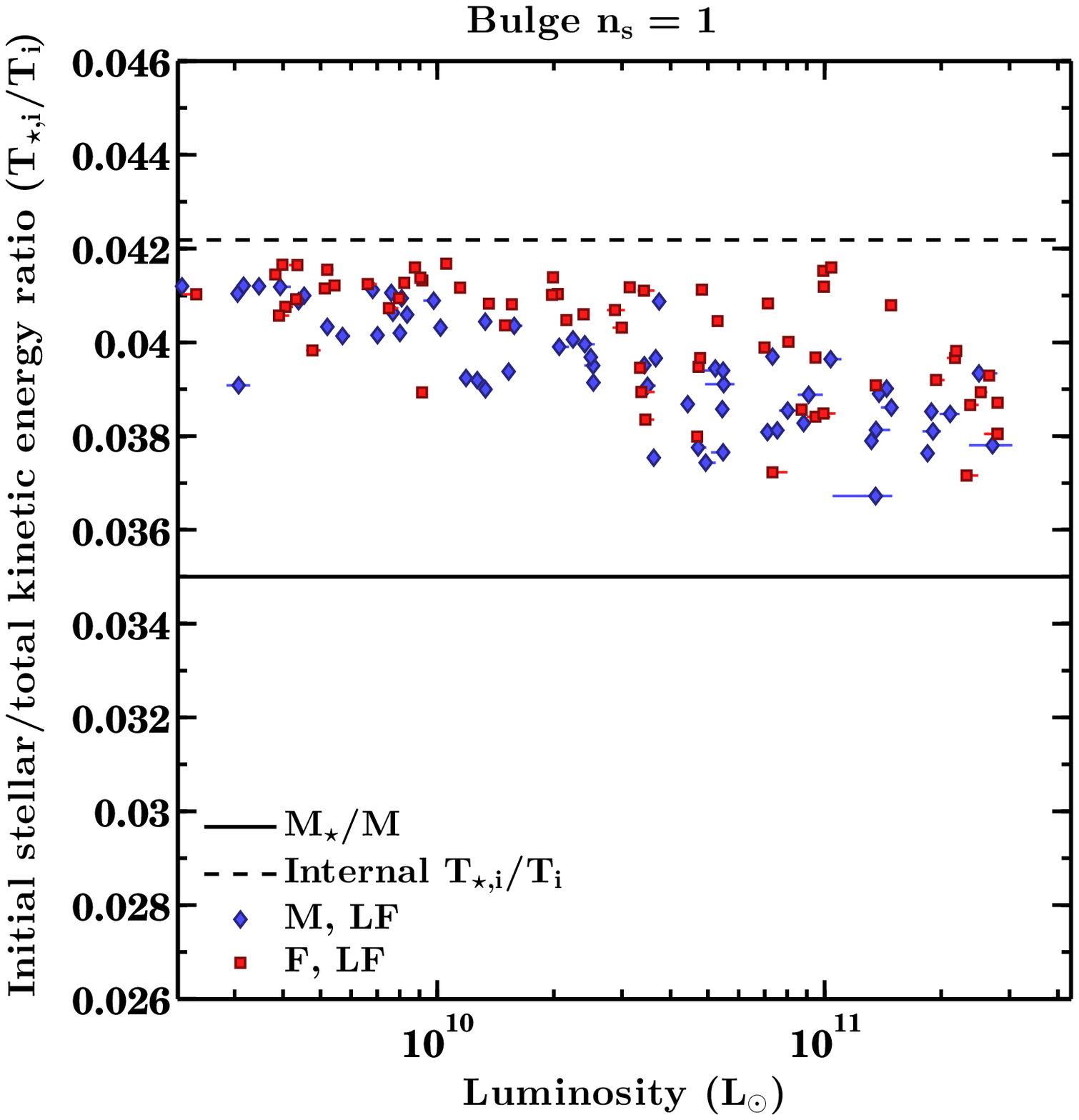}
\includegraphics[width=0.50\textwidth]{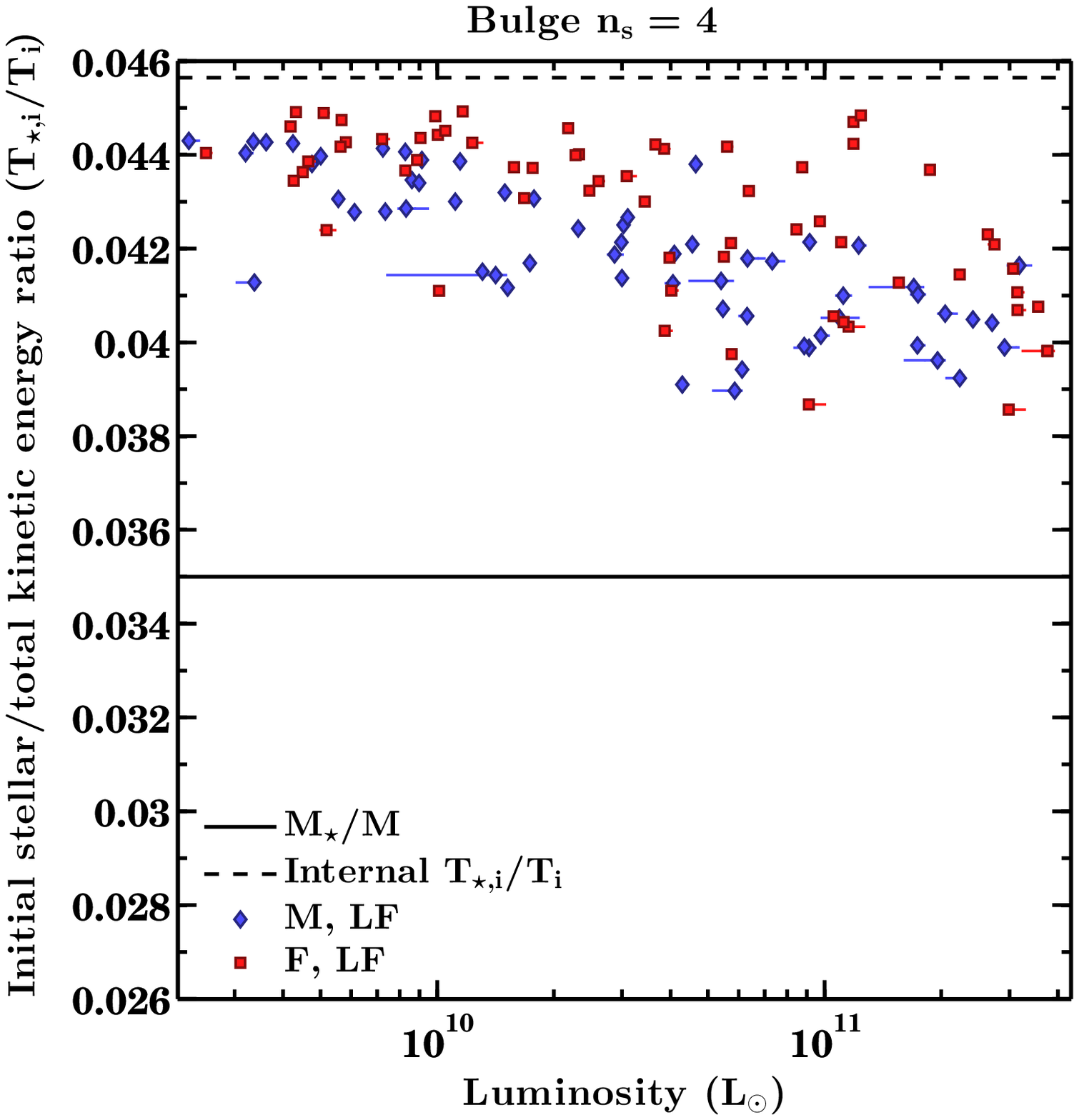}
\caption{Initial stellar kinetic energy ratios. The solid line shows the fraction of kinetic energy in stars for isolated galaxies (which are all self-similar); the points show the totals including orbital energy. More massive groups are dominated by the kinetic energy of the galaxy orbits and lie closer to the stellar mass fraction of 0.035, since orbital energy is equally distributed amongst stars and dark matter.
\label{fig:ekin}}
\end{figure*}

\figref{ekin} shows the initial fraction of stellar-to-total kinetic energy, $T_{\star}/T$ - the main variable component of $q_{T}$. The fraction is constant for each individual galaxy, since they are re-scaled versions of each other, and only differs for the two bulge models (0.0422 for $n_{s}$=1 and 0.0456 for $n_{s}$=4). However, the \emph{orbital} kinetic energy is distributed equally between the stars and dark matter, and since the stellar mass fraction is initially about 0.035, the initial value of $T_{\star}/T$ has a minimum of 0.035 in groups where the orbital kinetic energy dominates over the internal energies of each galaxy. 

The fact that more massive groups are dominated by orbital energy and not the internal motions with galaxies is a consequence of the scaling relations imposed on galaxies and groups. The galaxies are scaled to follow a Tully-Fisher relation $V \propto M^{0.29}$ (see Paper I). The groups are scaled such that the density within the maximum radius is constant ($\rho$=constant, $R_{max} \propto M^{1/3})$), and so $V_{orbital} \propto M^{2/3}$ - a steeper scaling than the Tully-Fisher relation. This effectively imposes a non-homology from the initial conditions of the simulation, and one which is not necessarily present by construction in binary merger simulations.

\begin{figure}
\includegraphics[width=0.49\textwidth]{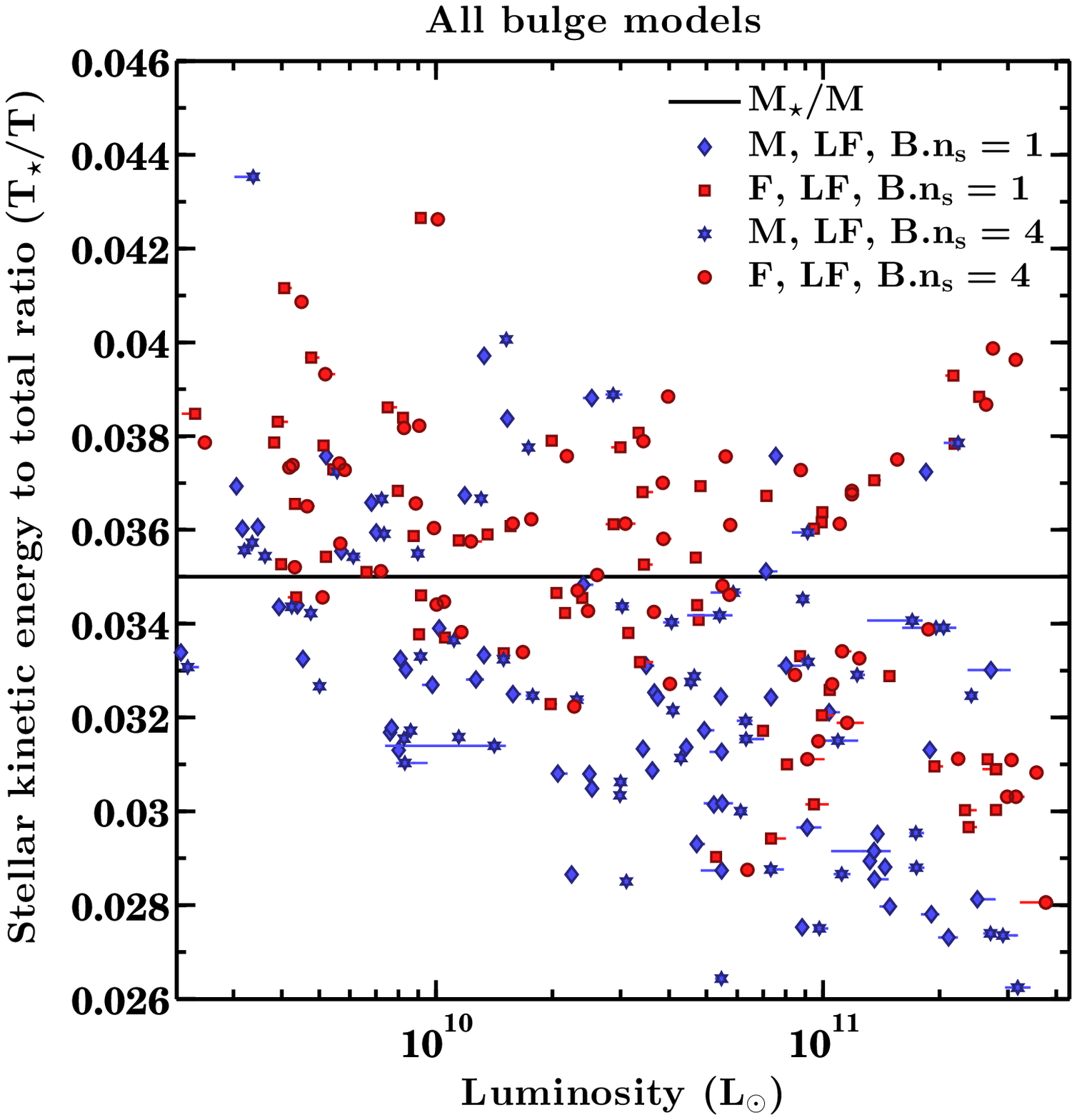}
\caption{Final stellar kinetic energy ratios after 10 Gyr, for comparison to \figref{ekin}.
\label{fig:ekinfin}}
\end{figure}

However, not all of the non-homology measured in the final remnants comes from the initial conditions. \figref{ekinfin} shows that the stellar kinetic energy ratio continues to drop in almost all of the galaxies - i.e., that the dark matter gains proportionally more kinetic energy than the stars. Furthermore, the stellar kinetic energy fraction drops more in the most massive groups. Interestingly, while all of the galaxies begin with larger $T_{\star}/T$ than $M_{\star}/M$ - i.e., a larger specific kinetic energy in stars than dark matter - a small majority end with $T_{\star}/T < M_{\star}/M$. This is not unrealistic - after all, most elliptical galaxies should have smaller $\sigma_{*}$ than $\sigma_{DM}$, since even the most massive ellipticals do not have dispersions larger than about 400 $\mathrm{km\ s^{-1}}$, whereas the most massive host halos in cosmological simulations can have 1000 $\mathrm{km\ s^{-1}}$ dispersions. It is likely that central group and cluster galaxies have $\sigma_{*}<\sigma_{DM}$; the same may not necessarily be true of satellite galaxies in subhalos. It is important that dissipationless merging can convert spirals with large stellar specific kinetic energies into ellipticals with lower specific kinetic energies than their halos. This is possible because the merging galaxies follow different scaling relations from the groups themselves, allowing large groups to be dominated by the orbital energies of whole galaxies, rather than of the orbits of masses within galaxies.

The fractional drop in $T_{\star}/T$ is about the same regardless of bulge type. The $B.n_{s}=4$ sample begins with a slightly stronger trend, simply because the $n_{s}=4$ bulge is more centrally concentrated and has a larger velocity dispersion at the same mass. Thus, the tilt is sensitive to the structure of the progenitor galaxies, and the larger tilt observed in the $B.n_{s}=4$ sample (\tabref{fpfits}) is at least partly because the initial spirals began with a larger bulge velocity dispersion at fixed mass. Nonetheless, the initial dependence of $T_{\star}/T$ on group mass or luminosity only steepens with extra merging, and the final ratios are not very sensitive to the initial values, so the initial conditions are only part of the reason why $q_{T}$ varies with galaxy and group mass.

The weakness of this formulation of the tilt - beyond that $T_{\star}/T$ is difficult to measure - is that $T_{\star}/T$ does not really return the observed FP to the virial FP the way that tilt terms like $M_{\star}/M$ do. Furthermore, the variation of $T_{\star}/T$ does not explain the trend in $M_{\star}/M$ itself. Nonetheless, it does illustrate how non-homology can originate from plausible merger histories in groups.

\section{Discussion}
\label{sec:discussion}

It is clear from the results of \subsecref{fp} that multiple dry mergers of spiral galaxies can produce a tilted FP, even if the merging galaxies are self-similar. This contrasts with the claim of \cite{HopCoxHer08} that ``dissipation appears to be both necessary and sufficient to explain the FP tilt''. Moreover, the merger remnants show exceptionally tight scaling relations, including just 0.02 dex scatter in the FP. This demonstrates that multiple dry merging is a promising channel for the formation of elliptical galaxies, and one that is certainly plausible for the formation of the most massive ellipticals. However, as we already showed in Paper I, producing less luminous, disky and fast-rotating ellipticals remains a challenge. S0s probably cannot be formed without some dissipation, if they are formed through mergers at all.

One of the major challenges to the merger theory of ellipticals - whether dissipational or not - comes from directly measured or inferred dark matter fractions in real ellipticals. As shown in \figref{fracdm_proj}, the dark matter fractions within $R_{e}$ are quite large - at least 40\% in projection (30\% in 3D) and up to 80 -- 90\%. These are not inconsistent with results from the Sloan Lens ACS survey (SLACS), who found similarly large dark matter fractions \citep{AugTreBol10} with a strong luminosity-dependent trend \citep{BarCzoKoo11}, assuming a Chabrier IMF. However, the situation is not as clear if the IMF is not universal. In particular, if the IMF is more ``bottom-heavy'' - i.e. closer to \citet{Sal55} - then these larger observed dark matter fractions could be over-estimated, and the simulated fractions would be too large. In some low-mass ellipticals a Salpeter IMF is nearly ruled out, as it implies a negative dark matter fraction, but near-zero dark matter fractions are not impossible. If only the most massive ellipticals have an IMF even more ``bottom-heavy'' or steeper than \citet{Sal55}, then some of the dark mass within $R_{e}$ could simply be faint M dwarfs, making the dark matter fraction substantially lower. The net effect from such a variable IMF would be to flatten the dark matter fraction relation in \figref{fracdm_proj}, which would shrink the tilt from the $M/M_{\star}$ term. This would be a major problem for dry and wet mergers alike, since most previous studies have found the dark matter fraction term to be a major contributor to the tilt.

The large dark matter fractions in dissipationless mergers would also appear to be at odds with results from dynamical modelling of nearby galaxies \cite{CapScoAla13}, which favor values closer to 20\% than 50\%. Such dynamical models have yet to be applied to merger simulations - both the Jeans and Schwarzschild modelling methods have only been tested against idealized galaxies \citep{LabCapEms12} but not triaxial merger remnants. Nonetheless, the low inferred dark matter fractions would seem to support dissipational merging, as such mergers can shrink $R_{e}$ and thus also lower $M/M_{\star}$ within $R_{e}$ significantly \citep{HopCoxHer08}. However, this requires the majority of ellipticals to be formed from relatively gas-rich major mergers. It is unclear whether gas fractions are really so high beyond $z>1$ \citep{NarBotDav12}, even in the most massive disks. Cosmological simulations also suggest that major mergers are less important for the formation of massive ellipticals than late minor mergers \citep{NaaJohOst09,OseNaaOst12}. \citet{RobCoxHer06} reproduced virtually the entire tilt of the stellar mass FP with \~40\% gas fractions, while our simulations suggest that dissipationless merging can account for half or more of the tilt. Thus, if multiple, dissipationless mergers do occur, then gas-rich mergers must be less common to avoid producing an excessively large tilt in the stellar mass FP, or gas fractions should be lower overall. It is not clear if this would still result in low dark matter fractions.

It is reassuring that the scatter in the simulation FP of just 0.02 dex is considerably smaller than the observed scatter, although it may not be smaller than the intrinsic scatter, which is difficult to estimate. Evidently, projection effects do not greatly enhance the FP scatter, a finding that is consistent with results from mergers of spheroidal galaxies in clusters \citep{NipStiCio03}. However, the simulation FP scatter should be taken as a lower limit, as it is partly a byproduct of the assumption of a zero-scatter Tully-Fisher relation for the progenitor spirals. As discussed in Paper I, there is significant scatter in the observed Tully-Fisher relation \citep[about 0.12 dex:][]{CouDutvdB07}, although the intrinsic scatter could be much smaller. If the \citet{HopCoxHer08} estimate that 0.1 dex scatter in the Tully-Fisher relation contributes about 0.04 dex scatter in the FP is correct, then the addition of 0.02 dex scatter in quadrature from the simulations would leave the simulation FP scatter consistent with observed values of 0.05--0.06 dex.

Although the non-zero tilt of the FP is a robust result of this work, the exact value of the tilt is sensitive to many factors. In \subsecref{fp}, we showed that the degree of tilt depends on the structure of the merging galaxies, such that merging galaxies with more concentrated bulges produces a steeper tilt. The merger history may also have an impact. While the raw number of mergers may not affect the tilt much, as long as it is larger than one, the details of the merger history do have an impact on the tilt, as \appref{randics} shows. The observed tilt could also vary if $M_{\star}/M_{DM}$ was a strong function of group mass rather than constant, as we have assumed, and if the bulge fractions and profiles of the merging galaxies were mass-dependent. It remains to be shown whether the exact values of the tilt of the FP emerge naturally from self-consistent cosmological merging, given all of the possible causes for a tilt in the FP - only some of which have been explored in this paper.

A number of possible interpretations exist for why and how a tilted FP is generated from multiple dissipationless mergers. In \subsecref{fp}, we showed that the majority of the tilt is caused by a variable dark matter fraction, and in turn by the fact that $M_{DM,R_{e},3D} \propto R_{e}^2$, whereas $M_{\star,R_{e},3D} \propto R_{e}^{1.72}$. These scalings clearly result in a dark matter fraction that increases with radius, although it is not obvious why these specific scalings are generated by mergers in groups. Nonetheless, such mass-dependent dark matter fractions are consistent both with results from simulations of binary mergers of gas-rich spirals \citep[e.g.][]{RobCoxHer06}, as well those from mergers of spheroidal galaxies in rich clusters \citep[e.g.][]{RusSpr09}.

In \secref{alttilt}, we introduced an alternative formulation of the tilt of the FP. In this formulation, the tilt is partly the result of the redistribution of kinetic and potential energy. One possible tilt term examined in \secref{origin} is the ratio of stellar to total kinetic energy, $T_{\star}/T$. $T_{\star}/T$ is mass-dependent, with more luminous galaxies having lower values. This difference is partly embedded in the initial conditions. More massive groups have a larger fraction of the kinetic energy in the orbital energies of galaxies within the group, rather than in internal motions within individual galaxies. The difference is also enhanced by the merging process, such that many galaxies end up with lower specific stellar kinetic energies ($T_{\star}/T < M_{\star}/M$), even though each progenitor galaxy began with $T_{\star}/T > M_{\star}/M$. Although $T_{\star}/T$ alone is not strictly the cause of the tilt of the FP, since it needs to be combined with potential energy terms, this formulation does help to explain why multiple mergers in groups are fundamentally different from the standard self-similar binary merger scenario.

The significant contribution of the $M_{\star}/L$ term to the tilt remains another major challenge for elliptical galaxy formation theories, and we are unaware of any self-consistent theories for its origin. \citet{RobCoxHer06} compared their stellar mass FP to near-infrared observations. A fully consistent theory for the formation of ellipticals should explain stellar population variations in visible bands as well. In principle, this could originate from dissipationless merging, if more massive ellipticals are formed from mergers of more massive galaxies with older/more metal-rich stellar populations with larger $M_{\star}/L$, but it is unknown if this could generate a steep enough tilt.

In \subsecref{mdyn}, we showed that the virial FP (\eqnref{virpfit}) defined by $M_{R_{e},3D}$, $R_{e}$ and $S$ (or $\sigma_{e}$) can be used to derive a dynamical mass, $M_{dyn}=M_{R_{e},3D}=cS^2R_{e}/G$, with $c\approx2.6$ and 0.06 dex scatter. We also examined alternative mass estimators proposed by \citet{WolMarBul10} ($M_{1/2}$) and \citet{CapScoAla13} ($M_{model}$). Of these various dynamical mass estimators, $M_{R_{e},3D}$ is the only one that can both be accurately estimated from projected quantities and derived from a near-virial FP fit. $M_{1/2}$ can be estimated from $r_{1/2}$, the 3D half-light radius; however, $r_{1/2}$ cannot be directly measured, and the approximation $r_{1/2}=1.34R_{e}$ for pure Sersic profiles does not hold for our galaxies, where $r_{1/2}\approx1.88R_{e}$. 

We also find that $M_{model}$ is best reproduced with a constant $c$=5.45 rather than $c$=4, and that $M_{model}$ does not define an exact virial FP, although these results may partly be due to systematic differences in $R_{e}$. $M_{model}$ can estimated fairly accurately from a near-virial FP using $R_{e,maj}$ and $\sigma$, rather than $\sigma_{e}$ or $S$; still, we suggest the use of $M_{R_{e},3D}$, as it is a physical mass within a well-defined radius. It is a curious coincidence that all three mass estimators are reasonably well fit by $c\approx$ 2.6--2.7 (if using $M_{model}$/2, and particular size and kinematic tracers). Ultimately, though, all of these dynamical masses only trace the total mass or dark matter fraction within $R_{e}$, while they underestimate the total mass of the galaxy by at least an order of magnitude.

As a final speculative note, we have limited discussion of the systematics in profile fitting, opting to match methodologies between simulations and observations instead. This is not to say that single Sersic fits are an ideal choice. Some systematic effects from Sersic fits to SDSS-quality images were noted in Paper I, and there is abundant discussion of observational systematics in the literature \citep[e.g.][]{MeeVikBer13}. More generally, the fundamental plane does not exist just because $R_{e}$ is a ``special'' radius. In principle, any size or luminosity measure should generate a fundamental plane, likely with slightly different tilt. As an example, one could measure fundamental planes with $2R_{e}$, $R_{e}/2$, or non-parametric $R_{50}$, $R_{25}$, $R_{10}$, etc. Just as ratios like $R_{90}/R_{50}$ are used as a proxy for concentration, other combinations of sizes would reflect differences in the surface brightness profiles of galaxies, much like the $n_{s}$ parameter attempts to do. In fact, any theory of elliptical galaxy formation should produce galaxies following \emph{every} observed fundamental plane relation, regardless of the definition of the size, velocity and mass or luminosity parameter.

\section{Conclusions}
\label{sec:conclusions}

Using collisionless simulations of mergers of spiral galaxies in groups, we have investigated whether the central merger remnants follow a similar fundamental plane relation to observed ellipticals. The following points summarize the conclusions:
\begin{enumerate}
\item Dissipationless mergers of multiple spiral galaxies in groups produce remnants resembling elliptical galaxies. These remnants lie on a tight fundamental plane (FP) relation with $a \approx 1.7$ and $b \approx 0.3$, which is tilted relative to the virial FP in the same sense as the observed FP.
\item The tilt from collisionless mergers could be responsible for about a third to half of the full observed FP tilt, explaining most of the variation in the $b$ parameter but only part of the change in $a$.
\item The simulation tilt is closer still to the observed stellar mass FP tilt and could explain most of the tilt that is not attributable to stellar population variations. Mergers of galaxies with more concentrated bulges and with fully randomized orbits both produce a more significant tilt.
\item The primary contributor to the simulation tilt is the variable dark matter fraction within $R_{e}$. Structural non-homology from a variable virial parameter $k$ may also contribute a small amount, but this may be mainly due to systematics in extracting $R_{e}$ from single Sersic profile fits.
\item Since \emph{multiple} collisionless mergers can produce a tilted FP, dissipation is not strictly necessary to create a tilted FP. Some dissipation is likely needed to increase central densities, shrink $R_{e}$ and raise $\sigma$. However, the 40\% gas fractions quoted by \cite{RobCoxHer06} as being necessary to reproduce the FP may be an overestimate.
%\item The virial FP is not strictly an extension of the scalar virial theorem, because the virial ratios of biased subsets of particles in collisionless systems (such as stars or the most bound particles within $R_{e},3D$) are not generally unity, or even exactly constant.
\item Although the virial FP does not strictly follow from the scalar virial theorem, the combination of $R_{e}$, $M_{R_{e},3D}$, and $S$ (or $\sigma_{e}$) do define almost precisely a virial FP with minimal scatter. As a result, the virial FP can be used to estimate a dynamical mass $M_{dyn}=M_{R_{e},3D}=cS^2R_{e}/G$, with $c\approx2.6$ and 0.06 dex scatter. We find that $M_{R_{e},3D}$ is the only true, dimensionally correct dynamical mass of the various mass estimators tested.
\end{enumerate}

\section{Acknowledgments}
\label{sec:acknowledgments}

D.T. would like to thank R. Abraham, F. van den Bosch, M. Cappellari, P. Nair, and T. Mendel for fruitful discussions and for providing data used herein. D.T. acknowledges the support of Ontario Graduate Scholarships for this work. J.D. and H.Y. acknowledge support from grants from the National Science Engineering Research Council of Canada. H.Y. acknowledges support from the Canada Research Chair program. Computations were performed on the gpc supercomputer at the SciNet HPC Consortium \citep{LokGruGro10}. SciNet is funded by: the Canada Foundation for Innovation under the auspices of Compute Canada; the Government of Ontario; Ontario Research Fund - Research Excellence; and the University of Toronto. Additional computations were performed at the Canadian Institute for Theoretical Astrophysics.

\appendix

\section{Sensitivity to Initial Conditions}
\label{app:randics}

The initial conditions in the first two subsamples were correlated, in the sense that groups of similar mass had similar positions of satellite galaxies, but randomized orbits (see Paper I for details). A third subsample had entirely randomized positions and orbits for each galaxy. This was designed to test how sensitive the tilt is to orbital configurations in each group.

\begin{table}
\centering
\caption{Sersic model Fundamental Plane Fits of Different Samples}
Simulations: Ten equally-spaced projections, randomly oriented \\
\begin{tabular}{ccccccc}
\hline 
B.$\mathrm{n_{s}}$ & Subsample & $a$ & $b$ & Intercept & R.M.S. \\
\hline
1 & 1 & 1.77 & 0.34 & -10.04 $\pm$ 0.10 & 0.02 \\
1 & 2 & 1.82 & 0.34 & -10.19 $\pm$ 0.10 & 0.02 \\
1 & 3 & 1.69 & 0.30 & -9.01 $\pm$ 0.09 & 0.03 \\
\hline
4 & 1 & 1.65 & 0.29 & -8.77 $\pm$ 0.06 & 0.02 \\
4 & 2 & 1.68 & 0.29 & -8.89 $\pm$ 0.07 & 0.02 \\
4 & 3 & 1.61 & 0.28 & -8.41 $\pm$ 0.05 & 0.02 \\
\hline
\end{tabular}
\tablecomments{Fundamental plane fits to subsamples with less (1,2) and more (3) randomized initial conditions; other table definition as in \tabref{fpfits}. All fits are unweighted. Fully random orbits (subsample 3) produce slightly larger tilts, especially for B.$\mathrm{n_{s}}=1$. Errors on $a$ are uniformly 0.01 and errors on $b$ are smaller than 0.005.
\label{tab:fpic}}
\end{table}

\tabref{fpic} lists the tilt for all three subsamples. Subsample 3 with completely random orbits shows the largest tilt of the three, whereas subsamples 1 and 2 are largely consistent with one another. This suggests that the tilt is somewhat sensitive to both the placement of galaxies within the group and their initial velocities/orbits. Moreover, the difference between B.$\mathrm{n_{s}}=1$ and B.$\mathrm{n_{s}}=4$ is much smaller in subsample 3, so there is some uncertainty in the relative importance of initial galaxy structure to the tilt.

\bibliography{taranu}

\end{document}